\documentclass[twocolumn]{aastex63}

\usepackage{amssymb}	
\usepackage{amsmath}	
\usepackage{mathrsfs} %
\usepackage{array}
\usepackage{bm}
\usepackage{CJK}
\usepackage{graphicx}	

\usepackage{xifthen}
\usepackage{xcolor}
\usepackage{xspace}
\usepackage{natbib}
\usepackage{comment}
\usepackage{xcolor}

\usepackage{hyperref}

\begin{document}
\begin{CJK*}{UTF8}{gbsn}
\title{Estimating JUNO's sensitivity to solar neutrino-to-antineutrino conversion\\ and neutrino magnetic moments}
\author{C. V. Ventura}
\author{Saul J. Panibra Churata}
\affiliation{Department of General Science, Universidad Continental,\\
Av. Los Incas 04002, Arequipa, Per\'u}
\begin{abstract}
We investigated JUNO's sensitivity to a possible conversion of solar neutrinos into antineutrinos via the spin-flavor precession (SFP) mechanism, and assessed the implications for constraining the neutrino-magnetic moment (NMM). Using a sensitivity-based framework appropriate for counting experiments with no prior observations, we derive 90\% C.L.\ ensemble-average sensitivities on the solar antineutrino flux for 1.8--16.8 MeV and 8.0--16.8 MeV. 
For the entire energy window, the results do not improve the restrictions of other experiments; the relevance occurs in the highes-energy window. In this window, we report a flux of $\phi_{\mathrm{lim}}\le 4.01\times10^{1}\ \mathrm{cm^{-2}\,s^{-1}}$ and a probability of $P_{\nu_e\rightarrow\bar{\nu}_e}\le 2.07\times10^{-5}$, the latter normalized to the ${}^8$B flux above threshold, $\Phi_{\rm SSM}(E>8~\mathrm{MeV})$.
Assuming transverse solar magnetic fields of $B_\perp=50$ and $100$~kG, the corresponding magnetic-moment sensitivities are $\mu_\nu\le 7.27\times10^{-11}\,\mu_B$ and $3.64\times10^{-11}\,\mu_B$ in the high-energy window.  These results highlight that JUNO has the potential to achieve sensitivities comparable to the most stringent astrophysical limits; in particular, the high-energy selection (8.0--16.8~MeV) provides a sensitivity that is competitive with current results, while the full-energy window remains primarily limited by near-reactor backgrounds.
\end{abstract}

\section{Introduction}
The investigation of a possible unknown source of antineutrinos to explain observed events unrelated to known-origin sources is crucial for understanding the nature of the neutrino. A hypothetical flux of solar antineutrinos cannot be ruled out as a potential explanation for this unknown source.

Research confirms that neutrinos have mass, a property that leads to the phenomenon of neutrino oscillations, where neutrinos can change flavor as they propagate. This also implies they may possess a small magnetic moment capable of interacting with intense magnetic fields~\citep[e.g.,][]{cisneros1971effect}, such as the solar magnetic field. This interaction induces SFP in Majorana neutrinos, causing a flavor oscillation that converts a neutrino into an antineutrino, $\nu\rightarrow\bar{\nu}$~\citep{giunti2009neutrino,Giunti:2014ixa}. Therefore, weak antineutrino fluxes cannot be excluded from the observations.

Over the past decade, neutrino experiments like Borexino~\citep{bellini2011study}, KamLAND~\citep{Eguchi:2003gg}, and Super-Kamiokande~\citep{fukuda1998evidence} have significantly contributed to setting upper limits on exotic neutrino signals. Borexino, renowned for its sensitivity to low-energy neutrinos, has placed stringent constraints on the solar antineutrino flux and the possible magnitude of the NMM. KamLAND, with its capability to detect antineutrinos from both geophysical and nuclear reactor sources, has also provided valuable detection limits. Meanwhile, Super-Kamiokande, focused on detecting high-energy solar and atmospheric neutrinos, has imposed key restrictions on flavor oscillations and neutrino conversion under exotic mechanisms such as SFP in solar magnetic fields.

Recent studies have further refined constraints on the NMM (\(\mu_\nu\)). 
The most restrictive astrophysical bounds now arise from recent analyses of stellar-cooling in red-giant-branch (TRGB) stars, rather than from earlier globular-cluster estimates. 
In particular, using \textit{Gaia}-based geometric calibrations of $\omega$ Centauri and NGC 4258,~\citet{PhysRevD.102.083007} derived upper limits of 
\(\mu_\nu < 1.2 \times 10^{-12} \mu_B\) and \(\mu_\nu < 1.5 \times 10^{-12} \mu_B\) (95\% C.L.), respectively, 
representing the most stringent astrophysical constraints to date. 
Earlier analyses of globular-cluster stars, where excessive neutrino emission would accelerate stellar cooling, 
had set a weaker limit of \(\mu_\nu < 3 \times 10^{-12} \mu_B\)~\citep{raffelt1992non}.

Direct experimental constraints mainly originate from reactor neutrino experiments. 
The GEMMA experiment at the Kalinin Nuclear Power Plant currently holds the most stringent direct bound, 
\(\mu_{\bar{\nu}_e} < 3.2 \times 10^{-11} \mu_B\) at 90\% confidence level (C.L.), using a high-purity germanium detector positioned 13.9 meters from the reactor core~\citep{beda2010upper}. 
Some studies have proposed that incorporating atomic ionization effects in neutrino-electron scattering analyses could enhance this limit, lowering it to 
\(\mu_{\bar{\nu}_e} < 5 \times 10^{-12} \mu_B\)~\citep{wong2010enhanced}. 
However, this claim remains debated, as alternative theoretical evaluations suggest that atomic effects do not significantly influence the neutrino-electron cross-section within the relevant energy range~\citep{voloshin2010neutrino}.

Meanwhile, Borexino has set an independent constraint on the effective NMM by analyzing the recoil-electron spectrum from solar neutrinos, yielding 
\(\mu_\nu < 5.4 \times 10^{-11} \mu_B\) at 90\% C.L.~\citep{bellini2011study}. 
Additionally, cosmological constraints from Big Bang Nucleosynthesis (BBN) and the Cosmic Microwave Background (CMB) impose a competitive limit, excluding values above 
\(\mu_\nu < 1.6 \times 10^{-11} \mu_B\)~\citep{carenza2024strong}. 
Cosmological observations of the effective number of relativistic species (\(N_{\text{eff}}\)) 
from BBN and the CMB disfavor Dirac, flavor-universal, diagonal NMM exceeding 
\(2.7 \times 10^{-12}\,\mu_B\), yielding constraints that surpass current XENONnT/LZ limits 
and are comparable to those from stellar-cooling analyses~\citep{lishaoping}.

In massive stars, nonzero NMM at the \( (2{-}4)\times10^{-11} \mu_B\) level can qualitatively alter late burning stages and mass thresholds, 
even enabling a thermonuclear CO (carbon-oxygen) explosion within a massive star (Type I.7) scenario~\citep{Heger_2009}. 
Future developments in reactor neutrino experiments and next-generation cosmological surveys are expected to refine these constraints.

\section{The JUNO Experiment and Detector Characteristics}
The Jiangmen Underground Neutrino Observatory (JUNO) is a large-scale neutrino experiment currently under construction in China, designed primarily to determine the neutrino mass hierarchy and measure oscillation parameters with unprecedented precision. The central detector consists of a spherical liquid scintillator target with a diameter of 35.4 meters, containing 20 kton of ultra-pure liquid scintillator, making it the largest detector of its kind ever built. Surrounding the scintillator, an array of 17,612 20-inch photomultiplier tubes (PMTs) and 25,600 3-inch PMTs provides an optical coverage exceeding 75\%, enabling an exceptional energy resolution of \(\sim3\%\) at $1$ MeV, crucial for distinguishing the oscillation patterns of reactor neutrinos at a baseline of 53 km. The detector is located \(\sim700\) meters underground to shield against cosmic ray backgrounds, significantly enhancing its sensitivity to low-energy neutrino interactions~\citep[see,][]{an2016neutrino,stock2024status}.

Beyond its primary goal of reactor neutrino oscillation studies, JUNO will serve as a multi-purpose neutrino observatory, capable of detecting solar neutrinos, geoneutrinos, atmospheric neutrinos, and supernova neutrinos. Thanks to its large volume and the high purity of the liquid scintillator, JUNO will be able to detect Boron-8 solar neutrinos with unprecedented precision, allowing the search for potential signals of solar neutrino-to-antineutrino conversion induced by SFP in solar magnetic fields. Furthermore, the detector's external veto system, consisting of a water Cherenkov muon detector and an array of plastic scintillators, will help suppress cosmic muon backgrounds, further improving its ability to study neutrino physics and probe physics beyond the Standard Model~\citep[see,][for more details]{an2016neutrino,stock2024status}.

\section{Framework}\label{environment}

\subsection{Sensitivity prescription}

In a counting experiment where data have not yet been collected, the appropriate figure of merit is the \textit{median upper limit} (or sensitivity), rather than the ``maximum allowed signal'' $S_{\mathrm{lim}}$ corresponding to a particular outcome, which is only relevant once measurements are available.

For $b\lesssim 15$ we use the exact~\citet[Table XII]{Feldman:1997qc} (FC) median limits, whereas for $b\gtrsim 16$ we adopt the one-sided Gaussian approximation,
\begin{equation}
\mu^{90}_{\rm UL}=1.28\sqrt{b},\label{eq:gaussian_aproximation}
\end{equation}
as recommended by the~\citet{PDG2024Stats}. This reproduces the FC construction with sub-percent accuracy once $b\gtrsim 16$.

This background dependent approach ensures correct coverage for all $b$ values while avoiding the computational cost of generating extended Feldman-Cousins confidence belts for the $\mathcal{O}(10^3)$ background events expected in the high-threshold analysis window. 

The resulting sensitivities---summarized in Table~\ref{tab:wideHigh} for the two prompt-energy ranges (1.8--16.8~MeV and 8--16.8~MeV)---are then translated into limits on the solar $\bar{\nu}_e$ flux, the spin-flavor conversion probability $P_{\nu_e\rightarrow\bar{\nu}_e}$, and the NMM ($\mu_\nu$), using the corresponding energy-averaged inverse beta decay cross sections, $\sigma_A = 3.40\times 10^{-42} \ \mathrm{cm}^2$ and $\sigma_B = 6.64\times 10^{-42} \ \mathrm{cm}^2$.  

\subsection{Upper Flux Limit and Conversion Probability}

The upper limit on the solar antineutrino flux~\citep{bellini2011study}, \(\phi_{\mathrm{lim}}\), can be determined according to:  
\begin{equation}
\phi_{\mathrm{lim}} = \frac{\mu^{90}_{\mathrm{UL}}}{\overline{\sigma} \, T \, n_{p} \, \epsilon} \quad [\mathrm{cm}^{-2}\,\mathrm{s}^{-1}],
\label{4.6}
\end{equation}  
where \(\mu^{90}_{\mathrm{UL}}\) is the sensitivity, representing the number of events not associated with any known source. The quantity \(\overline{\sigma}\) denotes the \textit{effective inverse beta decay} (IBD) cross section, averaged over the prompt-energy range of interest and weighted by the \(^8\)B solar neutrino spectrum. 

We obtain the flux-averaged IBD cross section in each window by weighting the energy-dependent cross section with the ${}^8$B spectral shape from \citet{Bahcall:1996qv} and averaging over the window limits (1.8--16.8 MeV and 8.0--16.8 MeV). The result reflects the relative contribution of each neutrino energy to the event yield in that window; efficiencies are taken as constant, so they factor out of the average.

The exposure time is \(T = 1.5768 \times 10^{8} \ \mathrm{s}\) (corresponding to five years of continuous data-taking), the number of free proton targets is \(n_p = 1.44 \times 10^{33}\)~\citep{Abrahao:2015rba}, and the overall detection efficiency is taken to be \(\epsilon \approx 1\).

The IBD threshold at $E_\nu\simeq 1.80$~MeV (hereafter 1.8~MeV) marks the minimum neutrino energy for the reaction $\bar{\nu}_e+p\to e^++n$, which is the primary detection channel in large liquid-scintillator detectors. For our selections, the window-averaged IBD cross sections are $\overline{\sigma}=3.40\times10^{-42}\ \mathrm{cm}^2$ in the $1.8$--$16.8$~MeV window and $\overline{\sigma}=6.64\times10^{-42}\ \mathrm{cm}^2$ in the $8.0$--$16.8$~MeV window, obtained using the \citeauthor{Strumia:2003zx} prescription folded over the analysis windows. Reactor $\bar{\nu}_e$ spectra are largely confined below 8~MeV and fall steeply above this energy; nevertheless, at JUNO baselines and reactor powers, the residual reactor contribution above 8~MeV is subdominant but non-negligible relative to cosmogenic and atmospheric backgrounds, and is therefore included explicitly in our background model (see Table~\ref{tab:wideHigh}).

\begin{table*}[t]
\centering
\caption{90\% C.L. sensitivities (ensemble-average upper limits, $\langle\mu^{90}_{\rm UL}\rangle$) for each
background category and for the total background in the two $E_\nu$ windows for JUNO (20 kt, 5 yr, $\epsilon=1$).
Quoted $b$ are the expected event counts; limits use the one-sided Gaussian approximation
$\mu^{90}_{\rm UL}=1.28\sqrt{b}$ for $b>0$ and the Feldman--Cousins sensitivity $\mu^{90}_{\rm UL}=2.44$ for $b=0$.
Flux limits are $\Phi_{\rm lim}=\mu^{90}_{\rm UL}/(\bar\sigma\,T\,n_p)$ with
$\bar\sigma_A T n_p = 0.772$ for 1.8--16.8 MeV and $\bar\sigma_B T n_p = 1.506$ for 8--16.8 MeV.
Probability limits are $P_{\nu_e\to\bar\nu_e}=\Phi_{\rm lim}/\Phi_{\rm SSM}(E>E_{\rm th})$ with
$\Phi_{\rm SSM}(E>1.8~\mathrm{MeV})=5.74\times10^6$ and $\Phi_{\rm SSM}(E>8.0~\mathrm{MeV})=1.94\times10^6\ \mathrm{cm}^{-2}\mathrm{s}^{-1}$.
\emph{Note:} “Far reactors” denotes regional cores (Daya Bay, Huizhou), while “World reactors” denotes the rest-of-world contribution; the two are disjoint.
}
\label{tab:wideHigh}
\small
\setlength{\tabcolsep}{12pt}
\begin{tabular}{lcccc}
\hline\hline
\multicolumn{5}{c}{\textbf{Energy window: 1.8--16.8 MeV}\quad($\bar{\sigma}_A T n_p = 0.772$)}\\
\hline
Background & $b$ (evt) & $\mu_{\mathrm{UL}}^{90}$ (evt) & $\Phi_{\mathrm{lim}}\;[\mathrm{cm}^{-2}\mathrm{s}^{-1}]$ &
$P_{\nu_e\to\bar{\nu}_e}$ \\
\hline
Near reactors        &  98671 & 402.1 & $5.21\times10^{2}$ & $9.07\times10^{-5}$ \\
Far reactors         &  11205 & 135.5 & $1.76\times10^{2}$ & $3.07\times10^{-5}$ \\
Geoneutrinos         &    837 &  37.0 & $4.80\times10^{1}$ & $8.36\times10^{-6}$ \\
World reactors       &   1756 &  53.6 & $6.95\times10^{1}$ & $1.21\times10^{-5}$ \\
Accidentals          &    798 &  36.2 & $4.68\times10^{1}$ & $8.15\times10^{-6}$ \\
$^{9}$Li/$^{8}$He    &   2143 &  59.3 & $7.68\times10^{1}$ & $1.34\times10^{-5}$ \\
NC atmospheric       &    247 &  20.1 & $2.61\times10^{1}$ & $4.55\times10^{-6}$ \\
Fast neutrons        &    247 &  20.1 & $2.61\times10^{1}$ & $4.55\times10^{-6}$ \\
$^{13}$C($\alpha$,n) &     91 &  12.2 & $1.58\times10^{1}$ & $2.75\times10^{-6}$ \\
\hline
\textbf{Total A}     & \textbf{115\,995} & \textbf{435.9} &
\textbf{$5.65\times10^{2}$} & \textbf{$9.84\times10^{-5}$} \\
\hline\hline
\multicolumn{5}{c}{\textbf{Energy window: 8--16.8 MeV}\quad($\bar{\sigma}_B T n_p = 1.506$)}\\
\hline
Background & $b$ (evt) & $\mu_{\mathrm{UL}}^{90}$ (evt) & $\Phi_{\mathrm{lim}}\;[\mathrm{cm}^{-2}\mathrm{s}^{-1}]$ &
$P_{\nu_e\to\bar{\nu}_e}$ \\
\hline
Near reactors        &   712 & 34.15 & $2.27\times10^{1}$ & $1.17\times10^{-5}$ \\
Far reactors         &    52 &  9.23 & $6.13$             & $3.16\times10^{-6}$ \\
$^{9}$Li/$^{8}$He    &  1294 & 46.04 & $3.06\times10^{1}$ & $1.58\times10^{-5}$ \\
Fast neutrons        &   145 & 15.41 & $1.02\times10^{1}$ & $5.26\times10^{-6}$ \\
NC atmospheric       &   145 & 15.41 & $1.02\times10^{1}$ & $5.26\times10^{-6}$ \\
World reactors       &    20 &  5.72 & $3.80$             & $1.96\times10^{-6}$ \\
Geoneutrinos         &     0 &  2.44 & $1.62$             & $8.36\times10^{-7}$ \\
Accidentals          &     0 &  2.44 & $1.62$             & $8.36\times10^{-7}$ \\
$^{13}$C($\alpha$,n) &     0 &  2.44 & $1.62$             & $8.36\times10^{-7}$ \\
\hline
\textbf{Total B}     & \textbf{2\,223} & \textbf{60.35} &
\textbf{$4.01\times10^{1}$} & \textbf{$2.07\times10^{-5}$} \\
\hline
\end{tabular}
\end{table*}

\subsection{Conversion Probabilities and Magnetic Moments}\label{sec:3.3}
The probability of converting a neutrino into an antineutrino, \(P_{\nu \to \bar{\nu}}\)~\citep[see][for more detail]{Akhmedov:2002mf,barger2002imprint}, can be written as the ratio of the flux limit \(\phi_{\text{lim}}\) to the solar neutrino flux predicted by the Standard Solar Model (SSM), here \(\phi_{\text{SSM}} = 5.88 \times 10^6 \, \text{cm}^{-2}\text{s}^{-1}\) refers to the \emph{full} \({}^8\)B spectrum ~\citep{Serenelli:2009ww}:
\begin{equation}
P_{\nu \to \bar{\nu}} = \frac{\phi_{\text{lim}}}{\phi_{\text{SSM}}}.\label{q3}
\end{equation}

For the high-energy window (8.0--16.8 MeV), only a subset of the ${}^8$B spectrum lies above threshold. 
Accordingly, we evaluate the conversion probability using the SSM flux integrated above the analysis threshold, 
$\Phi_{\rm SSM}(E>E_{\rm th})$. With the spectral shape from \citeauthor{Bahcall:1996qv} normalized to the \citeauthor{Serenelli:2009ww} SSM flux, this gives 
$\Phi_{\rm SSM}(E>8.0~\mathrm{MeV}) = 1.94\times10^{6}\ \mathrm{cm^{-2}\,s^{-1}}$ for the high-energy selection, 
while for the full window we use $\Phi_{\rm SSM}(E>1.8~\mathrm{MeV}) = 5.74\times10^{6}\ \mathrm{cm^{-2}\,s^{-1}}$. 
Using the window-appropriate $\Phi_{\rm SSM}(E>E_{\rm th})$ prevents an artificial inflation of the inferred sensitivity 
at high threshold and is reflected in Table~\ref{tab:wideHigh}.

The probability of solar neutrino conversion into antineutrinos in the Large Mixed Angle (LMA) scenario is given by~\citet{Akhmedov:2002mf},
\begin{equation*}
P ( \nu \to \bar{\nu} ) \approx 1.8 \times 10^{-10} \sin^2 2\theta \left( \frac{\mu_{\bar{\nu}}}{10^{-12} \mu_B} \frac{B_{\perp}(0.05R_{\odot})}{10 \text{kG}} \right)^2
\end{equation*}

where $\mu_{\bar{\nu}}$ is the transition NMM and $B_{\perp}$ is the transverse component of the solar magnetic field in the neutrino production region. The value $0.05 R_{\odot}$ is considered because it corresponds to the region where \(^8\)B neutrinos are generated~\citep[see,][]{Bahcall:1996qv,Bahcall:2004fg,1963ApJ...137..344B}, which dominate the JUNO detection spectrum in the relevant energy range ( 8--16 MeV). This equation shows that the conversion of $\nu_e$ into  $\bar{\nu}_e$ inside the Sun is highly suppressed unless $\mu_{\bar{\nu}}B_{\perp}$ is significantly large. The non-detection of a flux of solar $\bar{\nu}_e$ imposes an upper limit on $P(\nu_e \to \bar{\nu}_e)$, constraining the NMM to

\begin{equation}
\mu_{\bar{\nu}} \leq 7.4 \times 10^{-7} \left( \frac{P_{\nu \to \bar{\nu}}}{\sin^2 2\theta_{12}} \right)^{1/2} \frac{\mu_B}{B_{\perp} [\text{kG}]}\label{q4}
\end{equation}

which represents one of the most stringent constraints on $\mu_{\bar{\nu}}$ obtained to date. In this equation, \(\sin^2 2\theta_{12} = 0.86\) is the neutrino mixing parameter \citep{gando2011constraints,barger2002imprint,aliani2003determination,de2002solar}. 

The conversion of solar neutrinos into antineutrinos via SFP critically depends on the transverse magnetic field in the solar interior. 

The NMM is a hypothetical property that would allow neutrinos to interact with magnetic fields~\citep{giunti2009neutrino}. In the Standard Model, the Dirac NMM is predicted to be extremely small, on the order of \(10^{-19}\mu_{B}\) (Bohr magneton), making it undetectable by current experimental means~\citep{lindner2017revisiting}. However, in theories beyond the Standard Model, such as Majorana neutrino models or supersymmetry, imposing a limit on the NMM could be sufficiently feasible in the JUNO experiment.
\section{Analysis and Results}

In JUNO, we do not measure the NMM directly. Instead, we derive the ensemble-average 90\,\% C.L.\ upper limit on the signal, $\mu^{90}_{\mathrm{UL}}$, and convert it into a sensitivity on the $\bar{\nu}_e$ flux.
Dividing this flux limit by the SSM $^{8}$B prediction yields the conversion probability $P_{\nu_e\to\bar{\nu}_e}$; inserting $P$ into the SFP relation then gives the corresponding limit on the transition NMM.
Any future observation of a magnetic moment substantially larger than this bound would imply efficient $\nu_e\!\to\!\bar{\nu}_e$ conversion in strong solar fields and would thus signal physics beyond the Standard Model.

The residual background rates used here are taken from \citet[Table~3]{Abusleme_2025} of the JUNO mass-ordering study. That work reports post-selection rates (events/day) for seven categories —geoneutrinos, distant reactor $\bar\nu_e$, accidentals, $^{9}$Li/$^{8}$He ($\beta$-n), fast neutrons, atmospheric NC interactions, and $^{13}$C($\alpha$,n)— together with their physical origin and spectral shapes. Each category mimics the IBD signature to some extent: some produce genuine $\bar\nu_e$ IBD events (reactors, geoneutrinos), while others yield prompt-delayed correlated pairs that are experimentally indistinguishable from IBD ($^{9}$Li/$^{8}$He, fast neutrons, NC). Accidentals, although uncorrelated, also enter the IBD window through random prompt-delay coincidences. Since our goal is to bound a potential solar $\bar\nu_e$ signal —which would be detected via IBD—all categories must be included in the expected background $b$~\citep[see,][also the discussion on background modelling and uncorrelated bin-to-bin uncertainties]{Abusleme_2025}.

In addition to these categories, we explicitly include the reactor background from the eight near cores located at $\simeq 52.5$ km (Taishan-1, Taishan-2, and Yangjiang-1$\ldots$6) as well as the far reactors (e.g., Daya Bay at $\sim 215$ km and Huizhou at $\sim 265$ km). These reactor contributions were evaluated with our in-house simulation that computes the expected event yield by integrating the differential rate $\,\mathrm{d}N/\mathrm{d}E_\nu$ over the two analysis windows~\citep{Abrahao:2015rba}. The code implements three-flavor vacuum oscillations ($P_{ee}$ with PDG parameters), the \citeauthor{Strumia:2003zx} IBD cross section, and fission spectra with typical fission fractions (0.59/0.28/0.07/0.06)~\citep{Huber:2011wv,Mueller.83.054615}. For each core we use its thermal power and baseline to build the flux, and then fold $\Phi\times\sigma_{\rm IBD}\times P_{ee}$ over energy, assuming a 20~kt fiducial mass, 5~yr live time, and $\epsilon_{\rm det}=1$. This procedure yields per-window reactor$+$geo counts that, when combined with the remaining categories from \citet{Abusleme_2025}$\,$(geoneutrinos, accidentals, $^{9}$Li/$^{8}$He, fast-$n$, $^{13}$C($\alpha$,n), NC), produce updated totals of $b_A=115\,995$ and $b_B=2\,223$ events. The corresponding Gaussian one-sided sensitivities are $\mu^{90}_{\rm UL}=1.28\sqrt{b}$, i.e., $435.9$ and $60.35$ events for the low- and high-energy windows, respectively. We adopt the JUNO \emph{normal mass-ordering} per-category rate uncertainties to build the window-level rate systematic $\delta b/b$ and propagate them, together with $1.6\%$ detector-scale systematics and the $11\%$ ${}^{8}$B flux uncertainty, into the error budget reported in Table~\ref{tab:errorBudget}.
Inverse-beta-decay events from power reactors are explicitly included, as they dominate the background for a solar $\bar\nu_e$ signal at JUNO; see Table \ref{tab:wideHigh} for representative reactor IBD rates. Unlike Borexino, where large reactor standoff suppresses this contribution, JUNO's proximity to multiple cores renders the reactor background the leading component in the analysis windows.

From these rates we compute the expected counts over 5~yr and derive the ensemble-average 90\% C.L. upper limit $\mu_{\mathrm{UL}}$. For $b\le 15$ we use the exact~\citet{Feldman:1997qc} construction; for $b\gtrsim16$ we adopt the one-sided Gaussian approximation $\mu_{\mathrm{UL}}^{90}=1.28\sqrt{b}$, whose coverage differs from the exact result by $<1\%$ once $b\gtrsim 16$ \citep[Table.~XII]{Feldman:1997qc}; see also \citet{PDG2024Stats}. The resulting $\mu_{\mathrm{UL}}$ values are then converted into limits on the $\bar\nu_e$ flux, the conversion probability $P_{\nu_e\to\bar\nu_e}$, and the magnetic moment $\mu_\nu$ using the window-dependent averaged IBD cross sections (Sec.~\ref{sec:3.3}).

We reconstructed the residual background spectra~\footnote{The analysis codes used in this work are publicly available at Zenodo: \url{https://doi.org/10.5281/zenodo.17613794}.} by digitizing Fig. 3 of \emph{Potential to identify neutrino mass ordering with reactor antineutrinos at JUNO} (JUNO-NMO) \citep{Abusleme_2025}, interpolating the shapes over 1–5 MeV, and extending them above 5 MeV with an exponential tail whose slope is fixed by the local log-slope at 5 MeV. The tail has no tunable parameters: its decay constant is set by the measured log-slope at 5~MeV, and we enforce physically motivated endpoints (e.g., geoneutrinos stop at $\sim 3.3$~MeV, $^{13}\mathrm{C}(\alpha,n)$ at $\sim 7$--$7.5$~MeV). Each category is normalized once---either to the JUNO--NMO post-selection rate in 0.8--12.0~MeV (Table~3) or to a validated 1--5~MeV integral---avoiding double anchoring. Because reactor IBDs are our background (not signal), we compute the near- and far-reactor contributions with an independent reactor model (\citeauthor{Huber:2011wv}-\citeauthor{Mueller.83.054615} spectra, three-flavor oscillations, \citeauthor{Strumia:2003zx} $\sigma_{\mathrm{IBD}}$, baseline/power). Finally, we integrate the normalized spectra over 5~yr in 1.8--16.8~MeV and 8.0--16.8~MeV to obtain $b_A$ and $b_B$\citep{Ventura2025Software}. This procedure reproduces the official totals where they apply and adds only the physically warranted high-energy part.

It is important to emphasize that only the total background in each energy window is physically meaningful for these limits: probabilities and fluxes are not additive across categories. Per-category numbers are shown solely for comparison and pedagogical purposes; the reported physics limits are those derived from the total $b$ in each window.

We evaluated the sensitivity to the NMM within the SFP framework using the ensemble-average limits derived in this work. The two energy windows considered are 1.8--16.8 MeV and 8--16.8 MeV. 

\begin{figure*}[ht!]
    \centering
    \includegraphics[width=0.9\textwidth]{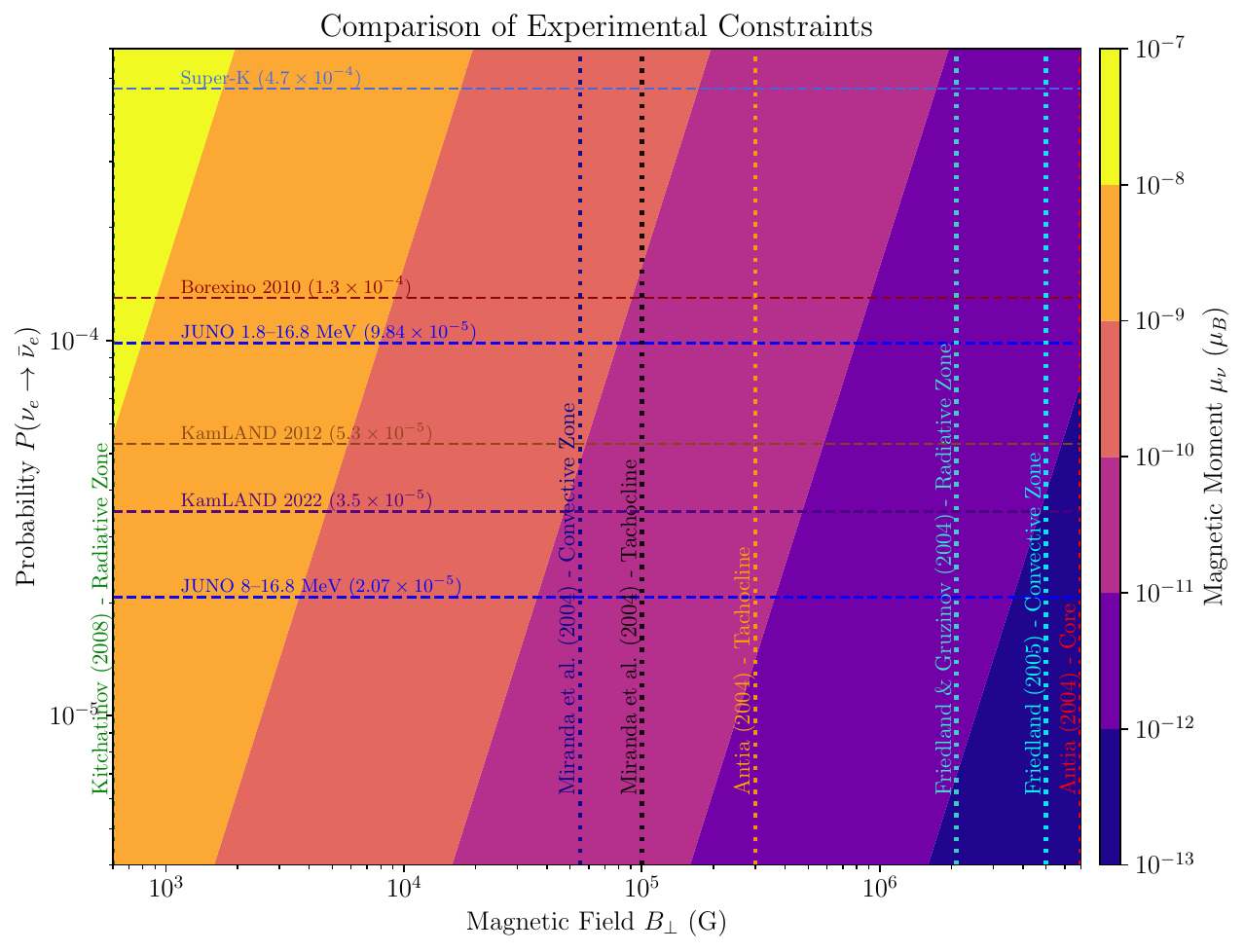} 
   \caption{Reference comparison of published upper limits on the conversion probability
$P(\nu_e\!\to\!\bar\nu_e)$ and their illustrative mapping to magnetic moment sensitivities.
\textbf{Important:} the experimental (and JUNO projected) limits on $P$ are field-independent;
the abscissa $B_\perp$ is used here only to visualize how a given probability would translate into
$\mu_\nu$ under the scaling $\mu_\nu = 7.4\times10^{-7}\,B_\perp^{-1}\sqrt{P/\sin^22\theta_{12}}$
with $\sin^22\theta_{12}=0.86$. The coloured band encodes the corresponding $\mu_\nu$ (in $\mu_B$) for
any pair $(P,B_\perp)$, and should not be interpreted as implying a dependence of $P$ on $B_\perp$.
Horizontal dashed lines show the 90\% C.L.\ limits on $P$ from Borexino (2010), KamLAND (2012, 2022),
Super-Kamiokande, and the JUNO ensemble-average sensitivities for the two prompt-energy windows
($1.8-16.8$ MeV and $8.0-16.8$ MeV). Vertical dotted lines mark benchmark field strengths associated
with different solar regions (convective zone, tachocline, radiative zone, core).
Plotted ranges: $600~\mathrm{G}\le B_\perp\le 7~\mathrm{MG}$ and $5\times10^{-6}\le P\le 6\times10^{-4}$.}
    \label{fig22}
\end{figure*}

The corresponding conversion probability sensitivities, which are independent of the solar field, are
$P_{\nu_e\to\bar\nu_e}\le 9.84\times10^{-5}$ for the $1.8$--$16.8$~MeV window and
$P_{\nu_e\to\bar\nu_e}\le 2.07\times10^{-5}$ for the $8$--$16.8$~MeV window.
To quantify the impact of these results, the flux sensitivities for the two ranges are
$\phi_{\bar\nu_e,\lim}\le 5.65\times10^{2}\ \mathrm{cm}^{-2}\,\mathrm{s}^{-1}$ and
$4.01\times10^{1}\ \mathrm{cm}^{-2}\,\mathrm{s}^{-1}$, respectively, corresponding to
window-level background totals of $b_A=115\,995$ and $b_B=2\,223$ (see Table~\ref{tab:wideHigh}). In the full 1.8--16.8 MeV window, near/far reactor \mbox{$\bar\nu_e$} dominate the background and JUNO’s projected probability sensitivity does not surpass the tightest existing results. In contrast, in the 8.0--16.8 MeV window the reactor spectrum is strongly suppressed and, once probabilities are computed using $\Phi_{\rm SSM}(E>E_{\rm th})$, JUNO becomes competitive—potentially leading (see Fig.~\ref{fig22} and Table~\ref{tab:published_plus_mu}).

Beyond the rate-based approach adopted here, a binned spectral analysis in energy would provide a more precise estimate of the sensitivity. Explicitly incorporating spectral information improves signal-to-background discrimination and can lead to tighter limits than those obtained from window-integrated counts.

Theoretical considerations regarding the solar magnetic field structure provide essential context for interpreting these limits. The conversion of solar neutrinos into antineutrinos via the SFP mechanism is strongly governed by the strength and spatial structure of the transverse magnetic field ($B_{\perp}$) across different regions of the Sun. Helioseismic studies suggest that in the solar core ($r < 0.2 R_{\odot}$) the magnetic field could reach values up to $7$~MG, with a gradual decrease through the radiative zone (RZ)~\citep[see,][]{antia2008seismic}. In the \(^8\)B neutrino production region ($r \approx 0.05 R_{\odot}$), where high-energy neutrinos are particularly sensitive to magnetic interactions, the helioseismic model of \cite{antia2008seismic} estimates a field of approximately $6.56$~MG, which is sufficient to allow significant conversion even in the presence of Mikheyev Smirnov Wolfenstein (MSW) suppression. However, stability analyses indicate that toroidal fields in the RZ should not exceed $\sim 600$~G to avoid inducing instabilities~\citep[see,][]{kitchatinov2008stability}. When $B_{\perp}$ in this region falls below 600~G, conversion becomes negligible~\citep{bellini2011study}.

The lower bound $B_\perp\simeq 600$\,G reflects the
magnetohydrodynamic (MHD) stability threshold for long-lived toroidal fields
in the RZ; stronger toroidal fields would be unstable and thus unlikely to persist over evolutionary timescales.

Beyond the RZ, the tachocline ($r \sim 0.7 R_{\odot}$), located at the interface between the radiative and convective zones, hosts magnetic fields estimated between $100$ and $300$~kG~\citep[see,][]{antia2008seismic,friedland2005solar}. Some studies have considered the possibility of even stronger fields in this region, up to $1$~MG, which could enhance the conversion probability~\citep{antia2008seismic}. Importantly, conversion in the tachocline is less affected by MSW suppression than in the core, and its steep magnetic field gradients make it a potentially favorable site for SFP-driven conversion~\citep[see,][for more detail]{Akhmedov:2002mf}.

Further outward, the convective zone (CZ) itself is characterized by turbulent and stochastic magnetic fields in the range of $50$ to $100$~kG~\citep[see,][]{miranda2004constraining,miranda2004enhanced,raffelt1992non}. Turbulent magnetic field models indicate that if $B_{\perp} < 100$~kG, the conversion probability is negligible, whereas for values approaching $1$~MG, the effect could be substantial~\citep{miranda2004constraining,miranda2004enhanced}. Fluctuations on spatial scales of 100--1000~km match the solar-neutrino oscillation length, enabling resonant conditions for the SFP mechanism~\citep{miranda2004constraining}. Moreover, turbulence induces loss of spin coherence, which can further amplify the conversion probability~\citep{miranda2004enhanced}. These considerations suggest that while a field $B_{\perp} > 1$~MG at $0.05 R_{\odot}$ could overcome oscillation suppression in the core, values above $300$~kG in the tachocline or sufficiently coherent turbulent fields in the CZ could also support efficient $\nu_{e} \to \bar{\nu}_{e}$ conversion.

 Building on this framework, for a transverse field \(B_\perp=600~\mathrm{G}\) we obtain \(\mu_\nu \le 1.33\times10^{-8}\,\mu_B\) (1.8--16.8 MeV) and \(\mu_\nu \le 6.07\times10^{-9}\,\mu_B\) (8--16.8 MeV). For \(B_\perp=7~\mathrm{MG}\) the limits become \(\mu_\nu \le 1.13\times10^{-12}\,\mu_B\) and \(\mu_\nu \le 5.18\times10^{-13}\,\mu_B\), respectively (see Table~\ref{tabla3}). These results follow directly from the field scaling \(\mu_\nu\propto 1/B_\perp\) applied to the probability limits above, and reflect the improved background conditions at higher visible energy threshold.
 

\begin{table}[t]
\centering
\caption{Projected 90\% C.L. sensitivities of JUNO to $P_{\nu_e\to\bar\nu_e}$ in two windows;
$\mu^{90}_{\rm UL}=1.28\sqrt{b}$ (or $2.44$ if $b\!\to\!0$). For the high-energy window we use
$\Phi_{\rm SSM}(E>E_{\rm th})$. Quoted $\mu_\nu$ assume $B_\perp=600~\mathrm{G}$ and $7~\mathrm{MG}$
(with comparisons made primarily in $P$, noting $\mu_\nu \propto \sqrt{P}/B_\perp$).}
\label{tabla3}
\small
\renewcommand{\arraystretch}{1.1}
\begin{tabular}{lcccc}
\hline\hline
{Energy} &
$\Phi_{\bar\nu_e,\text{lim}}$ & $P_{\nu_e\to\bar\nu_e}$ &
$\mu_\nu(600\,\text{G})$ & $\mu_\nu(7\,\text{MG})$ \\
(MeV) & $[\,\text{cm}^{-2}\text{s}^{-1}]$ & $[10^{-5}]$ & $[\,\mu_B]$ & $[\,\mu_B]$ \\
\hline\hline
1.8--16.8 & $5.65\times10^{2}$ & 9.84  & $1.33\times10^{-8}$ & $1.13\times10^{-12}$ \\
8.0--16.8 & $4.01\times10^{1}$ & 2.07  & $6.07\times10^{-9}$  & $5.18\times10^{-13}$ \\
\hline
\end{tabular}
\end{table}

Within the same setup, the projected sensitivities at the 90\% C.L.\ are as follows.

Full window ($1.8$--$16.8$ MeV): 
the conversion probability satisfies 
$P_{\nu_e\to\bar{\nu}_e}\le 9.84\times10^{-5}$, 
which corresponds to an antineutrino flux bound 
$\Phi_{\bar{\nu}_e}\le 5.65\times10^{2}\ \mathrm{cm}^{-2}\,\mathrm{s}^{-1}$ 
(see Table~\ref{tab:wideHigh}). 
Interpreting this probability in terms of the NMM  gives
$\mu_\nu\le 1.58\times10^{-10}\,\mu_B$ for $B_\perp=50~\mathrm{kG}$ and 
$\mu_\nu\le 7.91\times10^{-11}\,\mu_B$ for $B_\perp=100~\mathrm{kG}$ (see Table~\ref{tab:published_plus_mu}).

High-energy window ($8$--$16.8$ MeV): 
the conversion probability satisfies 
$P_{\nu_e\to\bar{\nu}_e}\le 2.07\times10^{-5}$, 
corresponding to 
$\Phi_{\bar{\nu}_e}\le 4.01\times10^{1}\ \mathrm{cm}^{-2}\,\mathrm{s}^{-1}$ (see Table~\ref{tab:wideHigh}). 
The associated magnetic-moment limits are
$\mu_\nu\le 7.27\times10^{-11}\,\mu_B$ for $B_\perp=50~\mathrm{kG}$ and 
$\mu_\nu\le 3.64\times10^{-11}\,\mu_B$ for $B_\perp=100~\mathrm{kG}$ (see Table~\ref{tab:published_plus_mu}).

\begin{table}[t]
\centering
\caption{Relative uncertainties (\%) propagated from background counts to antineutrino
flux, conversion probability, and NMM, now including near (8 cores at $\sim$52.5 km) and far/world reactor backgrounds.
The statistical component is $\sigma_{\mathrm{stat}}=1/\sqrt{b}$; rate and shape systematics are taken from Tab.~3 of~\citet{Abusleme_2025}.
A common detector factor $\delta k/k=1.6\%$ (0.4\,\% cross section~\citep{Strumia:2003zx}, 1\,\% efficiency, 0.3\,\% target protons~\citep{an2016neutrino}) is included in $\delta\Phi/\Phi$.
The ${}^{8}\mathrm{B}$ flux carries an $11\%$ uncertainty~\citep{Vinyoles_2017}, which dominates $\delta P/P$ and therefore $\delta\mu/\mu$.
Results correspond to a 5-yr exposure; total backgrounds are updated to $b_{A}=115{,}995$ and $b_{B}=2{,}223$, which imply
$\mu_{\mathrm{UL}}^{90}=435.9$ and $60.35$ events, respectively, using the 90\% C.L.\ one-sided Gaussian sensitivity $\mu_{\mathrm{UL}}^{90}=1.28\sqrt{b}$.
For probability limits we use $\Phi_{\rm SSM}(E>1.8~\mathrm{MeV})=5.74\times10^6$ and $\Phi_{\rm SSM}(E>8.0~\mathrm{MeV})=1.94\times10^6\ \mathrm{cm}^{-2}\mathrm{s}^{-1}$.}
\label{tab:errorBudget}
\small
\begin{tabular}{lcccc}
\hline\hline
\multicolumn{5}{c}{\textbf{Energy window: 1.8--16.8 MeV}}\\ \hline
Background & $\delta b/b$ & $\delta\Phi/\Phi$ & $\delta P/P$ & $\delta\mu/\mu$\\
\hline
Near reactors                      &  2.2 &  1.9 & 11.2 &  5.6 \\
Far/world reactors                 &  2.0 &  1.9 & 11.2 &  5.6 \\
Geoneutrinos                       & 30.0 & 15.1 & 18.7 &  9.3 \\
Accidentals                        &  1.0 &  1.7 & 11.1 &  5.6 \\
$^{9}$Li/$^{8}$He                  & 20.0 & 10.1 & 15.0 &  7.5 \\
Fast neutrons                      &100.0 & 50.0 & 51.2 & 25.6 \\
$^{13}$C($\alpha$,n)$^{16}$O       & 50.0 & 25.1 & 27.4 & 13.7 \\
NC atmospheric                     & 50.0 & 25.1 & 27.4 & 13.7 \\
\hline
\textbf{Total A}                   & \textbf{2.0} & \textbf{1.9} & \textbf{11.2} & \textbf{5.6}\\
\hline\hline
\multicolumn{5}{c}{\textbf{Energy window: 8--16.8 MeV}}\\ \hline
Background & $\delta b/b$ & $\delta\Phi/\Phi$ & $\delta P/P$ & $\delta\mu/\mu$\\
\hline
$^{9}$Li/$^{8}$He                  & 20.0 & 10.2 & 15.0 &  7.5 \\
Fast neutrons                      &100.0 & 50.0 & 51.2 & 25.6 \\
NC atmospheric                     & 50.0 & 25.1 & 27.4 & 13.7 \\
\hline
\textbf{Total B}                   & \textbf{11.1} & \textbf{5.9} & \textbf{12.5} & \textbf{6.2}\\
\hline
\end{tabular}
\end{table}

The estimated magnetic moment of solar neutrinos, using a conservative magnetic field and the SFP mechanism, shows values on the order of \(\sim 10^{-11} \mu_B\), depending on the magnetic field and conversion probabilities used. Comparing these values with the results obtained from various experiments, several key observations can be made.

A salient outcome of our comparison (Figure~\ref{fig22}, Table~\ref{tab:published_plus_mu}) is the clear
window dependence of JUNO’s reach. In the full 1.8--16.8~MeV selection, the sensitivity remains
background-limited by power-reactor $\bar\nu_e$ (eight near cores at $\sim\!52.5$~km plus the far/world component;
see Table~\ref{tab:wideHigh}), and does not surpass the tightest existing results
[$P_{\nu_e\to\bar\nu_e}\!\simeq\!9.84\times10^{-5}$ for JUNO]. In contrast, in the 8.0--16.8~MeV window the
reactor spectrum is strongly suppressed; adopting the appropriate normalization to the $^8$B flux above threshold,
$\Phi_{\rm SSM}(E>E_{\rm th})$---with $\Phi_{\rm SSM}(E>8~\mathrm{MeV})=1.94\times10^{6}\ \mathrm{cm^{-2}\,s^{-1}}$---JUNO
becomes competitive and slightly leading, with a projected $P_{\nu_e\to\bar\nu_e}\simeq2.07\times10^{-5}$
(compared to KamLAND’s $\sim3.5\times10^{-5}$). This high-energy selection therefore captures JUNO’s comparative
advantage for $P_{\nu_e\to\bar\nu_e}$ and, by scaling $\mu_\nu\propto\sqrt{P}/B_\perp$, yields a modest but
meaningful tightening of the corresponding NMM sensitivity.

For the KamLAND experiment, an upper limit for the magnetic moment of solar neutrinos was obtained, \( \mu_\nu \leq 3.5 \times 10^{-12} \mu_B \) (90\% C.L.), based on the observation of the conversion of \(^8\)B solar neutrinos to antineutrinos~\citep{gando2012search}. This limit is one of the strictest obtained in solar neutrino experiments through this conversion process. 

In Borexino, a limit for the magnetic moment of solar neutrinos of \( \mu_\nu \leq 1.4 \times 10^{-12} \mu_B \) was derived, using experimental data and a conversion probability of \( p_{\nu \to \bar{\nu}} = 1.3 \times 10^{-4} \)~\citep{bellini2011study}. 

The Super-Kamiokande experiment, for its part, established a limit of \( \mu_\nu \leq 3.6 \times 10^{-10} \mu_B \) based on 1496 days of solar neutrino data. This limit was improved to \( \mu_\nu \leq 1.1 \times 10^{-10} \mu_B \) when additional restrictions from neutrino oscillation experiments were considered~\citep{abe2022search}.

Finally, the analysis from SNO on solar neutrinos also established a limit on the magnetic moment of \( \mu_\nu \leq 1.4 \times 10^{-12} \mu_B \), based on the conversion of \(^8\)B solar neutrinos to antineutrinos~\citep{Aharmim:2004uf}.

\begin{deluxetable*}{lcccccc}[h]
\tablecaption{Published 90\% C.L. limits (no re-scaling of $P$ or $\phi$). Magnetic moment values are uniformly rescaled to $B_\perp=50,\,100$ kG for comparisons across experiments; primary comparisons in the text are given in terms of $P_{\nu_e\to\bar\nu}$. \emph{Note:} $\mu_\nu=\tfrac{7.4\times10^{-7}}{B_\perp[\mathrm{kG}]}\sqrt{P_{\rm lim}/\sin^22\theta_{12}}$ with $\sin^22\theta_{12}=0.86$. \label{tab:published_plus_mu}}
\tablewidth{0pt}
\tablehead{
\colhead{Experiment} & \colhead{Energy threshold} & $P_{\rm lim}$ & Flux $\phi_{\bar\nu}$ & $B_\perp=50$ kG  & $B_\perp=100$ kG \\
&(MeV)& & [$\mathrm{cm^{-2}\,s^{-1}}$] & ($\mu_B$) & ($\mu_B$)
}
\startdata
CTF~\citep{Balata:2006db}                         & $> 1.8$             & $1.9\times10^{-2}$  & $1.1\times10^{5}$   & $2.20\times10^{-9}$ & $1.10\times10^{-9}$ \\
SNO~\citep{Aharmim:2004uf}                        & $4.0$--$14.8$       & $8.1\times10^{-3}$  & $4.09\times10^{4}$  & $1.44\times10^{-9}$ & $7.18\times10^{-10}$ \\
Super--Kamiokande~\citep{abe2022search}           & $9.3$--$17.3$       & $4.7\times10^{-4}$  & $4.04\times10^{4}$  & $3.46\times10^{-10}$ & $1.73\times10^{-10}$ \\
KamLAND (2004)~\citep{Eguchi:2003gg}              & $8.3$--$14.8$       & $2.8\times10^{-4}$  & $1250$              & $2.67\times10^{-10}$ & $1.33\times10^{-10}$ \\
Borexino (2011; high thr.)~\citep{bellini2011study} & $>7.3$             & $1.7\times10^{-4}$  & $990$               & $2.08\times10^{-10}$ & $1.04\times10^{-10}$ \\
Borexino (2011; combined)~\citep{bellini2011study}  & $>1.8$             & $1.3\times10^{-4}$  & $760$               & $1.82\times10^{-10}$ & $9.10\times10^{-11}$ \\
KamLAND (2012)~\citep{gando2012search}            & $8.3$--$31.8$       & $5.3\times10^{-5}$  & ---                  & $1.16\times10^{-10}$ & $5.81\times10^{-11}$ \\
KamLAND (2022)~\citep{abe2022limits}              & $8.3$--$30.8$       & $3.5\times10^{-5}$  & $60$                & $9.44\times10^{-11}$ & $4.72\times10^{-11}$ \\
\textbf{JUNO (this work, $b=115\,995$)}           & $1.8$--$16.8$       & $\mathbf{9.84\times10^{-5}}$ & $565$  & $\mathbf{1.58\times10^{-10}}$ & $\mathbf{7.91\times10^{-11}}$ \\
\textbf{JUNO (this work, $b=2\,223$)}             & $8.0$--$16.8$       & $\mathbf{2.07\times10^{-5}}$ & $40.1$ & $\mathbf{7.27\times10^{-11}}$ & $\mathbf{3.64\times10^{-11}}$ \\
\enddata
\end{deluxetable*}

Figure~\ref{fig22} provides a comparative overview of published limits on the conversion probability
\(P(\nu_e\!\to\!\bar\nu_e)\). The color map encodes the corresponding magnetic-moment limit
\(\mu_\nu \propto B_\perp^{-1}\sqrt{P}\) for benchmark transverse fields \(B_\perp\), while vertical
guide lines mark representative solar-field benchmarks (convective zone, tachocline, radiative zone,
core). The plotted ranges are \(600~\mathrm{G}\le B_\perp\le 7~\mathrm{MG}\) and
\(5\times10^{-6}\le P\le 6\times10^{-4}\).
Dashed horizontal lines indicate the 90\% C.L.\ bounds from
Super-Kamiokande \((4.7\times10^{-4})\)~\citep{abe2022search},
Borexino \((1.3\times10^{-4})\)~\citep{bellini2011study}, and
KamLAND \((5.3\times10^{-5},\,3.5\times10^{-5})\)~\citep{gando2012search,abe2022limits}.
Two additional horizontal lines show JUNO’s ensemble-average sensitivities for
1.8--16.8~MeV and 8.0--16.8~MeV, at \(P\le 9.84\times10^{-5}\) and \(P\le 2.07\times10^{-5}\),
respectively. While the full-window sensitivity does not surpass the tightest existing limit,
the 8.0--16.8~MeV selection lies well below previous results and thus implies stronger \(\mu_\nu\)
bounds across the benchmark field range.

Table~\ref{tab:errorBudget} summarizes the relative uncertainties propagated from the
background expectations to the flux, conversion probability, and magnetic–moment sensitivities.
For the full window (1.8--16.8 MeV) we obtain
\(\delta\Phi/\Phi\simeq 1.9\%\), \(\delta P/P\simeq 11.2\%\), and \(\delta\mu/\mu\simeq 5.6\%\).
For the high-energy window (8.0--16.8 MeV) the corresponding values are
\(\delta\Phi/\Phi\simeq 5.9\%\), \(\delta P/P\simeq 12.5\%\), and \(\delta\mu/\mu\simeq 6.2\%\).
These figures already include the common detector normalization \((\delta k/k = 1.6\%)\) with
\(k=\bar{\sigma}\,T\,N_p\,\varepsilon\) (0.4\% IBD cross section, 1.0\% efficiency, 0.3\% target
protons) and the 11\% uncertainty on the solar \({}^8\)B flux. After these inputs, the residual
error budget is dominated by background modeling---geoneutrinos and \({}^9\)Li/\({}^8\)He in the full
window, and \({}^9\)Li/\({}^8\)He plus fast neutrons (with a subleading atmospheric--NC component)
in the high-energy window.
  
Total uncertainties are obtained by adding in quadrature uncorrelated rate and statistical contributions across backgrounds.  We stress that these fractional errors are independent of the assumed solar field \(B_\perp\) (which only rescales \(\mu_\nu\propto1/B_\perp\)), and they are a factor \(2\)-\(3\) smaller than the quoted fractional errors in Borexino and KamLAND~\citep{bellini2011study,gando2012search}, underscoring the robustness of JUNO's sensitivity.
To place these results in context, the flux, probabilities, and magnetic moment obtained in this study are compared with previous experimental constraints. In summary, these comparisons, shown in Table~\ref{tab:published_plus_mu}, highlight the potential of JUNO to refine current limits and contribute substantially to the global understanding of neutrino properties.
\section{Conclusions} 
We have investigated the possible conversion of solar neutrinos into antineutrinos via SFP and assessed its implications for the NMM.
Using a sensitivity-based framework appropriate for counting experiments with no prior observations in this channel, we derived ensemble-average 90\% C.L. sensitivities to the solar $\bar{\nu}_e$ flux in two prompt-energy ranges, 1.8--16.8~MeV and 8.0--16.8~MeV.

Our updated background model, summarized in Table~\ref{tab:wideHigh}, shows that antineutrinos from the eight nearest reactors at a baseline of $\simeq 52.5$~km dominate the low–energy region at JUNO, while contributions from distant/world reactors, geoneutrinos, cosmogenics, fast neutrons, $^{13}$C($\alpha$,n), accidentals, and atmospheric NC interactions are also included.
Above $8$~MeV the reactor spectrum is strongly suppressed, leaving cosmogenic and atmospheric sources as the leading backgrounds; these are explicitly accounted for in the high-energy sensitivity calculation, where probabilities are normalized to the fraction of the ${}^{8}$B flux above threshold, $\Phi_{\rm SSM}(E>E_{\rm th})$.

For the full 1.8--16.8~MeV window we obtain
$\phi_{\mathrm{lim}}\le 5.65\times10^{2}\ \mathrm{cm^{-2}\,s^{-1}}$ and
$P_{\nu_e\rightarrow\bar{\nu}_e}\le 9.84\times10^{-5}$.
For the high–energy window the sensitivities are
$\phi_{\mathrm{lim}}\le 4.01\times10^{1}\ \mathrm{cm^{-2}\,s^{-1}}$ and
$P_{\nu_e\rightarrow\bar{\nu}_e}\le 2.07\times10^{-5}$
(the latter using $\Phi_{\rm SSM}(E>E_{\rm th})$).
These results, summarized in Table~\ref{tab:published_plus_mu}, reflect the improved background conditions achieved above the visible-energy threshold.

Translating the (field-independent) probability sensitivities into magnetic–moment sensitivities with $\mu_\nu \propto \sqrt{P}/B_\perp$, we quote representative benchmarks.
For tachocline-scale reference fields, $B_\perp=50$ and $100$~kG, we find
$\mu_\nu \le 1.58\times10^{-10}\,\mu_B$ and $7.91\times10^{-11}\,\mu_B$ (full window), and
$\mu_\nu \le 7.27\times10^{-11}\,\mu_B$ and $3.64\times10^{-11}\,\mu_B$ (high-energy window), respectively.
For solar-interior benchmarks discussed in the text, $B_\perp=600$~G and $7$~MG, the analogous sensitivities are
$\mu_\nu \le 1.33\times10^{-8}\,\mu_B$ and $1.13\times10^{-12}\,\mu_B$ (full window), and
$\mu_\nu \le 6.07\times10^{-9}\,\mu_B$ and $5.18\times10^{-13}\,\mu_B$ (high-energy window).
We emphasize that cross-experiment comparisons are made primarily in terms of the field-independent conversion probability; the mapping to $\mu_\nu$ depends on the assumed $B_\perp$ and is shown for illustrative benchmarks.

Placing these results in context with previous studies (Figure~\ref{fig22}, Table~\ref{tab:published_plus_mu}) reveals a clear window dependence:
in the 1.8--16.8~MeV region JUNO remains background-limited by near/far reactors and does not surpass the tightest existing limits;
in contrast, in the 8.0--16.8~MeV window, where reactor contributions are strongly suppressed and the probability is normalized to $\Phi_{\rm SSM}(E>E_{\rm th})$, JUNO is competitive and slightly leading.
Numerically, our projected probability sensitivity $P_{\rm lim}\simeq 2.07\times10^{-5}$ improves upon the KamLAND\,2022 result ($\sim 3.5\times10^{-5}$) by a factor $\sim1.7$, which corresponds to a modest $\sim$20\% tightening in $\mu_\nu$ (given $\mu_\nu\propto\sqrt{P}$).

Overall, these findings highlight the scientific relevance of dedicated solar-antineutrino searches in large liquid-scintillator detectors.
Because the near-reactor component is the dominant background at JUNO, continued refinement of the background prediction $b$-together with optimized energy selections and, where possible, spectral-shape likelihoods-will further enhance sensitivity to $\nu_e\rightarrow\bar{\nu}_e$ transitions and to $\mu_\nu$, advancing our understanding of neutrino electromagnetic properties and solar astrophysics.


\bibliographystyle{aasjournal}
\bibliography{ref}

@article{fukuda1998evidence,
  title = {Evidence for Oscillation of Atmospheric Neutrinos},
  author = {Fukuda, Y. and Hayakawa, T. and Ichihara, E. and Inoue, K. and Ishihara, K. and Ishino, H. and Itow, Y. and Kajita, T. and Kameda, J. and Kasuga, S. and Kobayashi, K. and Kobayashi, Y. and Koshio, Y. and Miura, M. and Nakahata, M. and Nakayama, S. and Okada, A. and Okumura, K. and Sakurai, N. and Shiozawa, M. and Suzuki, Y. and Takeuchi, Y. and Totsuka, Y. and Yamada, S. and Earl, M. and Habig, A. and Kearns, E. and Messier, M. D. and Scholberg, K. and Stone, J. L. and Sulak, L. R. and Walter, C. W. and Goldhaber, M. and Barszczxak, T. and Casper, D. and Gajewski, W. and Halverson, P. G. and Hsu, J. and Kropp, W. R. and Price, L. R. and Reines, F. and Smy, M. and Sobel, H. W. and Vagins, M. R. and Ganezer, K. S. and Keig, W. E. and Ellsworth, R. W. and Tasaka, S. and Flanagan, J. W. and Kibayashi, A. and Learned, J. G. and Matsuno, S. and Stenger, V. J. and Takemori, D. and Ishii, T. and Kanzaki, J. and Kobayashi, T. and Mine, S. and Nakamura, K. and Nishikawa, K. and Oyama, Y. and Sakai, A. and Sakuda, M. and Sasaki, O. and Echigo, S. and Kohama, M. and Suzuki, A. T. and Haines, T. J. and Blaufuss, E. and Kim, B. K. and Sanford, R. and Svoboda, R. and Chen, M. L. and Conner, Z. and Goodman, J. A. and Sullivan, G. W. and Hill, J. and Jung, C. K. and Martens, K. and Mauger, C. and McGrew, C. and Sharkey, E. and Viren, B. and Yanagisawa, C. and Doki, W. and Miyano, K. and Okazawa, H. and Saji, C. and Takahata, M. and Nagashima, Y. and Takita, M. and Yamaguchi, T. and Yoshida, M. and Kim, S. B. and Etoh, M. and Fujita, K. and Hasegawa, A. and Hasegawa, T. and Hatakeyama, S. and Iwamoto, T. and Koga, M. and Maruyama, T. and Ogawa, H. and Shirai, J. and Suzuki, A. and Tsushima, F. and Koshiba, M. and Nemoto, M. and Nishijima, K. and Futagami, T. and Hayato, Y. and Kanaya, Y. and Kaneyuki, K. and Watanabe, Y. and Kielczewska, D. and Doyle, R. A. and George, J. S. and Stachyra, A. L. and Wai, L. L. and Wilkes, R. J. and Young, K. K.},
  collaboration = {Super-Kamiokande Collaboration},
  journal = {\prl},
  volume = {81},
  issue = {8},
  pages = {1562--1567},
  numpages = {0},
  year = {1998},
  month = {Aug},
  publisher = {American Physical Society},
  doi = {10.1103/PhysRevLett.81.1562},
  url = {https://link.aps.org/doi/10.1103/PhysRevLett.81.1562}
}

@article{Giunti:2014ixa,
  title = {Neutrino electromagnetic interactions: A window to new physics},
  author = {Giunti, Carlo and Studenikin, Alexander},
  journal = {Rev. Mod. Phys.},
  volume = {87},
  issue = {2},
  pages = {531--591},
  numpages = {61},
  year = {2015},
  month = {Jun},
  publisher = {American Physical Society},
  doi = {10.1103/RevModPhys.87.531},
  url = {https://link.aps.org/doi/10.1103/RevModPhys.87.531}
}

@article{cisneros1971effect,
  title={Effect of neutrino magnetic moment on solar neutrino observations},
  author={Cisneros, Arturo},
  journal={\apss},
  volume={10},
  number={1},
  pages={87--92},
  year={1971},
  publisher={Springer},
  doi={10.1007/BF00654607},
  url={https://doi.org/10.1007/BF00654607}

}

@article{voloshin2010neutrino,
  title = {Neutrino Scattering on Atomic Electrons in Searches for the Neutrino Magnetic Moment},
  author = {Voloshin, M. B.},
  journal = {\prl},
  volume = {105},
  issue = {20},
  pages = {201801},
  numpages = {4},
  year = {2010},
  month = {Nov},
  publisher = {American Physical Society},
  doi = {10.1103/PhysRevLett.105.201801},
  url = {https://link.aps.org/doi/10.1103/PhysRevLett.105.201801}
}

@article{bellini2011study,
title = {Study of solar and other unknown anti-neutrino fluxes with Borexino at LNGS},
journal = {Physics Letters B},
volume = {696},
number = {3},
pages = {191-196},
year = {2011},
issn = {0370-2693},
doi = {https://doi.org/10.1016/j.physletb.2010.12.030},
url = {https://www.sciencedirect.com/science/article/pii/S037026931001422X},
author = {G. Bellini and J. Benziger and S. Bonetti and M. {Buizza Avanzini} and B. Caccianiga and L. Cadonati and F. Calaprice and C. Carraro and A. Chavarria and A. Chepurnov and D. D'Angelo and S. Davini and A. Derbin and A. Etenko and K. Fomenko and D. Franco and C. Galbiati and S. Gazzana and C. Ghiano and M. Giammarchi and M. Goeger-Neff and A. Goretti and E. Guardincerri and S. Hardy and Aldo Ianni and Andrea Ianni and M. Joyce and V.V. Kobychev and D. Korablev and Y. Koshio and G. Korga and D. Kryn and M. Laubenstein and T. Lewke and E. Litvinovich and B. Loer and P. Lombardi and L. Ludhova and I. Machulin and S. Manecki and W. Maneschg and G. Manuzio and Q. Meindl and E. Meroni and L. Miramonti and M. Misiaszek and D. Montanari and V. Muratova and L. Oberauer and M. Obolensky and F. Ortica and M. Pallavicini and L. Papp and L. Perasso and S. Perasso and A. Pocar and R.S. Raghavan and G. Ranucci and A. Razeto and A. Re and P. Risso and A. Romani and D. Rountree and A. Sabelnikov and R. Saldanha and C. Salvo and S. Schönert and H. Simgen and M. Skorokhvatov and O. Smirnov and A. Sotnikov and S. Sukhotin and Y. Suvorov and R. Tartaglia and G. Testera and D. Vignaud and R.B. Vogelaar and F. {von Feilitzsch} and J. Winter and M. Wojcik and A. Wright and M. Wurm and J. Xu and O. Zaimidoroga and S. Zavatarelli and G. Zuzel},
}

@ARTICLE{1963ApJ...137..344B,
       author = {{Bahcall}, J.~N. and {Fowler}, William A. and {Iben}, Jr., I. and {Sears}, R.~L.},
        title = "{Solar Neutrino Flux.}",
      journal = {\apj},
         year = 1963,
        month = jan,
       volume = {137},
        pages = {344-346},
          doi = {10.1086/147513},
       adsurl = {https://ui.adsabs.harvard.edu/abs/1963ApJ...137..344B}
}

@article{Bahcall:2004fg,
  title = {What Do We (Not) Know Theoretically about Solar Neutrino Fluxes?},
  author = {Bahcall, John N. and Pinsonneault, M. H.},
  journal = {\prl},
  volume = {92},
  issue = {12},
  pages = {121301},
  numpages = {4},
  year = {2004},
  month = {Mar},
  publisher = {American Physical Society},
  doi = {10.1103/PhysRevLett.92.121301},
  url = {https://link.aps.org/doi/10.1103/PhysRevLett.92.121301}
}

@article{Bahcall:1996qv,
      author         = "Bahcall, John N. and Lisi, E. and Alburger, D. E. and De
                        Braeckeleer, L. and Freedman, S. J. and Napolitano, J.",
      title          = "{Standard neutrino spectrum from B-8 decay}",
      journal        = "\prc",
      volume         = "C54",
      year           = "1996",
      pages          = "411-422",
      doi            = "10.1103/PhysRevC.54.411",
      eprint         = "nucl-th/9601044",
      archivePrefix  = "arXiv",
      primaryClass   = "nucl-th",
      reportNumber   = "IASSNS-AST-96-28, IASSNS-AST-9547",
      SLACcitation   = "%%CITATION = NUCL-TH/9601044;%%"
}

@article{an2016neutrino,
doi = {10.1088/0954-3899/43/3/030401},
url = {https://dx.doi.org/10.1088/0954-3899/43/3/030401},
year = {2016},
month = {feb},
publisher = {IOP Publishing},
volume = {43},
number = {3},
pages = {030401},
author = {An, Fengpeng and An, Guangpeng and An, Qi and Antonelli, Vito and Baussan, Eric and Beacom, John and Bezrukov, Leonid and Blyth, Simon and Brugnera, Riccardo and Avanzini, Margherita Buizza and Busto, Jose and Cabrera, Anatael and Cai, Hao and Cai, Xiao and Cammi, Antonio and Cao, Guofu and Cao, Jun and Chang, Yun and Chen, Shaomin and Chen, Shenjian and Chen, Yixue and Chiesa, Davide and Clemenza, Massimiliano and Clerbaux, Barbara and Conrad, Janet and D’Angelo, Davide and Kerret, Hervé De and Deng, Zhi and Deng, Ziyan and Ding, Yayun and Djurcic, Zelimir and Dornic, Damien and Dracos, Marcos and Drapier, Olivier and Dusini, Stefano and Dye, Stephen and Enqvist, Timo and Fan, Donghua and Fang, Jian and Favart, Laurent and Ford, Richard and Göger-Neff, Marianne and Gan, Haonan and Garfagnini, Alberto and Giammarchi, Marco and Gonchar, Maxim and Gong, Guanghua and Gong, Hui and Gonin, Michel and Grassi, Marco and Grewing, Christian and Guan, Mengyun and Guarino, Vic and Guo, Gang and Guo, Wanlei and Guo, Xin-Heng and Hagner, Caren and Han, Ran and He, Miao and Heng, Yuekun and Hsiung, Yee and Hu, Jun and Hu, Shouyang and Hu, Tao and Huang, Hanxiong and Huang, Xingtao and Huo, Lei and Ioannisian, Ara and Jeitler, Manfred and Ji, Xiangdong and Jiang, Xiaoshan and Jollet, Cécile and Kang, Li and Karagounis, Michael and Kazarian, Narine and Krumshteyn, Zinovy and Kruth, Andre and Kuusiniemi, Pasi and Lachenmaier, Tobias and Leitner, Rupert and Li, Chao and Li, Jiaxing and Li, Weidong and Li, Weiguo and Li, Xiaomei and Li, Xiaonan and Li, Yi and Li, Yufeng and Li, Zhi-Bing and Liang, Hao and Lin, Guey-Lin and Lin, Tao and Lin, Yen-Hsun and Ling, Jiajie and Lippi, Ivano and Liu, Dawei and Liu, Hongbang and Liu, Hu and Liu, Jianglai and Liu, Jianli and Liu, Jinchang and Liu, Qian and Liu, Shubin and Liu, Shulin and Lombardi, Paolo and Long, Yongbing and Lu, Haoqi and Lu, Jiashu and Lu, Jingbin and Lu, Junguang and Lubsandorzhiev, Bayarto and Ludhova, Livia and Luo, Shu and Vladimir Lyashuk and Möllenberg, Randolph and Ma, Xubo and Mantovani, Fabio and Mao, Yajun and Mari, Stefano M and McDonough, William F and Meng, Guang and Meregaglia, Anselmo and Meroni, Emanuela and Mezzetto, Mauro and Miramonti, Lino and Thomas Mueller and Naumov, Dmitry and Oberauer, Lothar and Ochoa-Ricoux, Juan Pedro and Olshevskiy, Alexander and Ortica, Fausto and Paoloni, Alessandro and Peng, Haiping and Jen-Chieh Peng and Previtali, Ezio and Qi, Ming and Qian, Sen and Qian, Xin and Qian, Yongzhong and Qin, Zhonghua and Raffelt, Georg and Ranucci, Gioacchino and Ricci, Barbara and Robens, Markus and Romani, Aldo and Ruan, Xiangdong and Ruan, Xichao and Salamanna, Giuseppe and Shaevitz, Mike and Valery Sinev and Sirignano, Chiara and Sisti, Monica and Smirnov, Oleg and Soiron, Michael and Stahl, Achim and Stanco, Luca and Steinmann, Jochen and Sun, Xilei and Sun, Yongjie and Taichenachev, Dmitriy and Tang, Jian and Tkachev, Igor and Trzaska, Wladyslaw and Waasen, Stefan van and Volpe, Cristina and Vorobel, Vit and Votano, Lucia and Wang, Chung-Hsiang and Wang, Guoli and Wang, Hao and Wang, Meng and Wang, Ruiguang and Wang, Siguang and Wang, Wei and Wang, Yi and Wang, Yi and Wang, Yifang and Wang, Zhe and Wang, Zheng and Wang, Zhigang and Wang, Zhimin and Wei, Wei and Wen, Liangjian and Wiebusch, Christopher and Wonsak, Björn and Wu, Qun and Wulz, Claudia-Elisabeth and Wurm, Michael and Xi, Yufei and Xia, Dongmei and Xie, Yuguang and Zhi-zhong Xing and Xu, Jilei and Yan, Baojun and Yang, Changgen and Yang, Chaowen and Yang, Guang and Yang, Lei and Yang, Yifan and Yao, Yu and Yegin, Ugur and Yermia, Frédéric and You, Zhengyun and Yu, Boxiang and Yu, Chunxu and Yu, Zeyuan and Zavatarelli, Sandra and Zhan, Liang and Zhang, Chao and Zhang, Hong-Hao and Zhang, Jiawen and Zhang, Jingbo and Zhang, Qingmin and Zhang, Yu-Mei and Zhang, Zhenyu and Zhao, Zhenghua and Zheng, Yangheng and Zhong, Weili and Zhou, Guorong and Zhou, Jing and Zhou, Li and Zhou, Rong and Zhou, Shun and Zhou, Wenxiong and Zhou, Xiang and Zhou, Yeling and Zhou, Yufeng and Zou, Jiaheng},
title = {Neutrino physics with JUNO},
journal = {Journal of Physics G: Nuclear and Particle Physics},
}

@article{Strumia:2003zx,
title = {Precise quasielastic neutrino/nucleon cross-section},
journal = {Physics Letters B},
volume = {564},
number = {1},
pages = {42-54},
year = {2003},
issn = {0370-2693},
doi = {https://doi.org/10.1016/S0370-2693(03)00616-6},
url = {https://www.sciencedirect.com/science/article/pii/S0370269303006166},
author = {Alessandro Strumia and Francesco Vissani}
}

@article{Abrahao:2015rba,
      author         = {Abrahão, Thamys and Minakata, Hisakazu and 
                       Nunokawa, Hiroshi and Quiroga, Alexander A.},
      title          = {Constraint on Neutrino Decay with Medium- 
                       Baseline Reactor Neutrino Oscillation Experiments},
      journal        = {JHEP},
      volume         = {11},
      year           = {2015},
      pages          = {001},
      doi            = {10.1007/JHEP11(2015)001},
      eprint         = {1506.02314},
      archivePrefix  = {arXiv},
      primaryClass   = {hep-ph},
      reportNumber   = {INT-PUB-15-024},
      SLACcitation   = {%%CITATION = ARXIV:1506.02314;%%},
      url            = {https://doi.org/10.1007/JHEP11(2015)001}
}

@article{Feldman:1997qc,
  title = {Unified approach to the classical statistical analysis of small signals},
  author = {Feldman, Gary J. and Cousins, Robert D.},
  journal = {\prd},
  volume = {57},
  issue = {7},
  pages = {3873--3889},
  numpages = {0},
  year = {1998},
  month = {Apr},
  publisher = {American Physical Society},
  doi = {10.1103/PhysRevD.57.3873},
  url = {https://link.aps.org/doi/10.1103/PhysRevD.57.3873}
}

@ARTICLE{Serenelli:2009ww,
       author = {{Serenelli}, Aldo M.},
        title = "{New results on standard solar models}",
      journal = {\apss},
         year = 2010,
        month = jul,
       volume = {328},
       number = {1-2},
        pages = {13-21},
          doi = {10.1007/s10509-009-0174-8},
archivePrefix = {arXiv},
       eprint = {0910.3690},
 primaryClass = {astro-ph.SR},
       adsurl = {https://ui.adsabs.harvard.edu/abs/2010Ap&SS.328...13S}
}

@article{Akhmedov:2002mf,
title = {Solar neutrino oscillations and bounds on neutrino magnetic moment and solar magnetic field},
journal = {Physics Letters B},
volume = {553},
number = {1},
pages = {7-17},
year = {2003},
issn = {0370-2693},
doi = {https://doi.org/10.1016/S0370-2693(02)03182-9},
url = {https://www.sciencedirect.com/science/article/pii/S0370269302031829},
author = {E.Kh. Akhmedov and João Pulido}
}

@article{Balata:2006db,
      author         = {Balata, M. and others},
      title          = {{Search for electron antineutrino interactions with the
                        Borexino counting test facility at Gran Sasso}},
      collaboration  = {Borexino},
      journal        = {Eur. Phys. J.},
      volume         = {C47},
      year           = {2006},
      pages          = {21-30},
      doi            = {10.1140/epjc/s2006-02560-4},
      eprint         = {hep-ex/0602027},
      archivePrefix  = {arXiv},
      primaryClass   = {hep-ex},
      SLACcitation   = {%%CITATION = HEP-EX/0602027;%%},
      url            = {https://doi.org/10.1140/epjc/s2006-02560-4}
}

@article{Aharmim:2004uf,
  title = {Electron antineutrino search at the Sudbury Neutrino Observatory},
  author = {Aharmim, B. and Ahmed, S. N. and Beier, E. W. and Bellerive, A. and Biller, S. D. and Boger, J. and Boulay, M. G. and Bowles, T. J. and Brice, S. J. and Bullard, T. V. and Chan, Y. D. and Chen, M. and Chen, X. and Cleveland, B. T. and Cox, G. A. and Dai, X. and Dalnoki-Veress, F. and Doe, P. J. and Dosanjh, R. S. and Doucas, G. and Dragowsky, M. R. and Duba, C. A. and Duncan, F. A. and Dunford, M. and Dunmore, J. A. and Earle, E. D. and Elliott, S. R. and Evans, H. C. and Ewan, G. T. and Farine, J. and Fergani, H. and Fleurot, F. and Formaggio, J. A. and Fowler, M. M. and Frame, K. and Frati, W. and Fulsom, B. G. and Gagnon, N. and Graham, K. and Grant, D. R. and Hahn, R. L. and Hallin, A. L. and Hallman, E. D. and Hamer, A. S. and Handler, W. B. and Hargrove, C. K. and Harvey, P. J. and Hazama, R. and Heeger, K. M. and Heintzelman, W. J. and Heise, J. and Helmer, R. L. and Hemingway, R. J. and Hime, A. and Howe, M. A. and Jagam, P. and Jelley, N. A. and Klein, J. R. and Kormos, L. L. and Kos, M. S. and Kr\"uger, A. and Krauss, C. B. and Krumins, A. V. and Kutter, T. and Kyba, C. C. M. and Labranche, H. and Lange, R. and Law, J. and Lawson, I. T. and Lesko, K. T. and Leslie, J. R. and Levine, I. and Luoma, S. and MacLellan, R. and Majerus, S. and Mak, H. B. and Maneira, J. and Marino, A. D. and McCauley, N. and McDonald, A. B. and McGee, S. and McGregor, G. and Mifflin, C. and Miknaitis, K. K. S. and Miller, G. G. and Moffat, B. A. and Nally, C. W. and Neubauer, M. S. and Nickel, B. G. and Noble, A. J. and Norman, E. B. and Oblath, N. S. and Okada, C. E. and Ollerhead, R. W. and Orrell, J. L. and Oser, S. M. and Ouellet, C. and Peeters, S. J. M. and Poon, A. W. P. and Rielage, K. and Robertson, B. C. and Robertson, R. G. H. and Rollin, E. and Rosendahl, S. S. E. and Rusu, V. L. and Schwendener, M. H. and Simard, O. and Simpson, J. J. and Sims, C. J. and Sinclair, D. and Skensved, P. and Smith, M. W. E. and Starinsky, N. and Stokstad, R. G. and Stonehill, L. C. and Tafirout, R. and Takeuchi, Y. and Te\ifmmode \check{s}\else \v{s}\fi{}i\ifmmode \acute{c}\else \'{c}\fi{}, G. and Thomson, M. and Tsui, T. and Van Berg, R. and Van de Water, R. G. and Virtue, C. J. and Wall, B. L. and Waller, D. and Waltham, C. E. and Tseung, H. Wan Chan and Wark, D. L. and West, N. and Wilhelmy, J. B. and Wilkerson, J. F. and Wilson, J. R. and Wittich, P. and Wouters, J. M. and Yeh, M. and Zuber, K.},
  collaboration = {SNO Collaboration},
  journal = {\prd},
  volume = {70},
  issue = {9},
  pages = {093014},
  numpages = {8},
  year = {2004},
  month = {Nov},
  publisher = {American Physical Society},
  doi = {10.1103/PhysRevD.70.093014},
  url = {https://link.aps.org/doi/10.1103/PhysRevD.70.093014}
}

@article{gando2012search,
doi = {10.1088/0004-637X/745/2/193},
url = {https://dx.doi.org/10.1088/0004-637X/745/2/193},
year = {2012},
month = {jan},
publisher = {The American Astronomical Society},
volume = {745},
number = {2},
pages = {193},
author = {Gando, A. and Gando, Y. and Ichimura, K. and Ikeda, H. and Inoue, K. and Kibe, Y. and Kishimoto, Y. and Koga, M. and Minekawa, Y. and Mitsui, T. and Morikawa, T. and Nagai, N. and Nakajima, K. and Nakamura, K. and Narita, K. and Shimizu, I. and Shimizu, Y. and Shirai, J. and Suekane, F. and Suzuki, A. and Takahashi, H. and Takahashi, N. and Takemoto, Y. and Tamae, K. and Watanabe, H. and Xu, B. D. and Yabumoto, H. and Yoshida, H. and Yoshida, S. and Enomoto, S. and Kozlov, A. and Murayama, H. and Grant, C. and Keefer, G. and Piepke, A. and Banks, T. I. and Bloxham, T. and Detwiler, J. A. and Freedman, S. J. and Fujikawa, B. K. and Han, K. and Kadel, R. and O'Donnell, T. and Steiner, H. M. and Dwyer, D. A. and McKeown, R. D. and Zhang, C. and Berger, B. E. and Lane, C. E. and Maricic, J. and Miletic, T. and Batygov, M. and Learned, J. G. and Matsuno, S. and Sakai, M. and Horton-Smith, G. A. and Downum, K. E. and Gratta, G. and Efremenko, Y. and Kamyshkov, Y. and Perevozchikov, O. and Karwowski, H. J. and Markoff, D. M. and Tornow, W. and Heeger, K. M. and Decowski, M. P.},
title = {SEARCH FOR EXTRATERRESTRIAL ANTINEUTRINO SOURCES WITH THE KamLAND DETECTOR},
journal = {The Astrophysical Journal}
}

@article{gando2011constraints,
  title = {Constraints on ${\ensuremath{\theta}}_{13}$ from a three-flavor oscillation analysis of reactor antineutrinos at KamLAND},
  author = {Gando, A. and Gando, Y. and Ichimura, K. and Ikeda, H. and Inoue, K. and Kibe, Y. and Kishimoto, Y. and Koga, M. and Minekawa, Y. and Mitsui, T. and Morikawa, T. and Nagai, N. and Nakajima, K. and Nakamura, K. and Narita, K. and Shimizu, I. and Shimizu, Y. and Shirai, J. and Suekane, F. and Suzuki, A. and Takahashi, H. and Takahashi, N. and Takemoto, Y. and Tamae, K. and Watanabe, H. and Xu, B. D. and Yabumoto, H. and Yoshida, H. and Yoshida, S. and Enomoto, S. and Kozlov, A. and Murayama, H. and Grant, C. and Keefer, G. and Piepke, A. and Banks, T. I. and Bloxham, T. and Detwiler, J. A. and Freedman, S. J. and Fujikawa, B. K. and Han, K. and Kadel, R. and O'Donnell, T. and Steiner, H. M. and Dwyer, D. A. and McKeown, R. D. and Zhang, C. and Berger, B. E. and Lane, C. E. and Maricic, J. and Miletic, T. and Batygov, M. and Learned, J. G. and Matsuno, S. and Sakai, M. and Horton-Smith, G. A. and Downum, K. E. and Gratta, G. and Efremenko, Y. and Perevozchikov, O. and Karwowski, H. J. and Markoff, D. M. and Tornow, W. and Heeger, K. M. and Decowski, M. P.},
  collaboration = {The KamLAND Collaboration},
  journal = {\prd},
  volume = {83},
  issue = {5},
  pages = {052002},
  numpages = {11},
  year = {2011},
  month = {Mar},
  publisher = {American Physical Society},
  doi = {10.1103/PhysRevD.83.052002},
  url = {https://link.aps.org/doi/10.1103/PhysRevD.83.052002}
}

@article{Eguchi:2003gg,
  author = {Eguchi, K. and Enomoto, S. and Furuno, K. and Ikeda, H. and Ikeda, K. and Inoue, K. and Ishihara, K. and Iwamoto, T. and Kawashima, T. and Kishimoto, Y. and Koga, M. and Koseki, Y. and Maeda, T. and Mitsui, T. and Motoki, M. and Nakajima, K. and Ogawa, H. and Owada, K. and Piquemal, F. and Shimizu, I. and Shirai, J. and Suekane, F. and Suzuki, A. and Tada, K. and Tajima, O. and Takayama, T. and Tamae, K. and Watanabe, H. and Busenitz, J. and Djurcic, Z. and McKinny, K. and Mei, D-M. and Piepke, A. and Yakushev, E. and Berger, B. E. and Chan, Y. D. and Decowski, M. P. and Dwyer, D. A. and Freedman, S. J. and Fu, Y. and Fujikawa, B. K. and Goldman, J. and Heeger, K. M. and Lesko, K. T. and Luk, K.-B. and Murayama, H. and Nygren, D. R. and Okada, C. E. and Poon, A. W. P. and Steiner, H. M. and Winslow, L. A. and Horton-Smith, G. A. and Mauger, C. and McKeown, R. D. and Tipton, B. and Vogel, P. and Lane, C. E. and Miletic, T. and Gorham, P. W. and Guillian, G. and Learned, J. G. and Maricic, J. and Matsuno, S. and Pakvasa, S. and Dazeley, S. and Hatakeyama, S. and Svoboda, R. and Dieterle, B. D. and DiMauro, M. and Detwiler, J. and Gratta, G. and Ishii, K. and Tolich, N. and Uchida, Y. and Batygov, M. and Bugg, W. and Efremenko, Y. and Kamyshkov, Y. and Kozlov, A. and Nakamura, Y. and Gould, C. R. and Karwowski, H. J. and Markoff, D. M. and Messimore, J. A. and Nakamura, K. and Rohm, R. M. and Tornow, W. and Young, A. R. and Chen, M-J. and Wang, Y-F.},
  title = {High Sensitivity Search for ${\overline{\ensuremath{\nu}}}_{e}$'s from the Sun and Other Sources at KamLAND},
  collaboration = {KamLAND Collaboration},
  journal = {\prl},
  volume = {92},
  issue = {7},
  pages = {071301},
  numpages = {5},
  year = {2004},
  month = {Feb},
  publisher = {American Physical Society},
  doi = {10.1103/PhysRevLett.92.071301},
  url = {https://link.aps.org/doi/10.1103/PhysRevLett.92.071301}
}

@article{giunti2009neutrino,
  title={Neutrino electromagnetic properties},
  author={Giunti, Carlo and Studenikin, Alexander},
  journal={Physics of Atomic Nuclei},
  volume={72},
  pages={2089--2125},
  year={2009},
  ISSN={1562-692X},
  url={http://dx.doi.org/10.1134/S1063778809120126},
  doi={10.1134/s1063778809120126},
  publisher={Springer}
}

@article{lindner2017revisiting,
  title={Revisiting large neutrino magnetic moments},
  author={Lindner, Manfred and Radov{\v{c}}i{\'c}, Branimir and Welter, Johannes},
  journal={Journal of High Energy Physics},
  volume={2017},
  number={7},
  pages={1--15},
  year={2017},
  ISSN={1029-8479},
  doi = {10.1007/JHEP07(2017)139},
  url = {https://doi.org/10.1007/JHEP07%282017%29139},
  publisher={Springer}
}

@article{raffelt1992non,
  title={Non-standard neutrino interactions and the evolution of red giants},
  author={Raffelt, Georg and Weiss, Achim},
  journal={\aap},
  volume={264},
  number = {2},
  pages={536--546},
  year={1992},
  adsurl = {https://ui.adsabs.harvard.edu/abs/1992A&A...264..536R},
}

@article{beda2010upper,
  title={Upper limit on the neutrino magnetic moment from three years of data from the GEMMA spectrometer},
  author={Beda, AG and Brudanin, VB and Egorov, VG and Medvedev, DV and Pogosov, VS and Shirchenko, MV and Starostin, AS},
  journal={arXiv e-prints},
  year={2010},
  doi = {10.48550/arXiv.1005.2736},
  url={https://arxiv.org/abs/1005.2736},
  publisher={Citeseer}
}

@article{wong2010enhanced,
  title={Enhanced Sensitivities for the Searches of Neutrino Magnetic Moments<? format?> through Atomic Ionization},
  author={Wong, Henry T and Li, Hau-Bin and Lin, Shin-Ted},
  journal={\prl},
  volume={105},
  number={6},
  pages={061801},
  year={2010},
  doi = {10.1103/PhysRevLett.105.061801},
  url = {https://doi.org/10.1103/PhysRevLett.105.061801},
  publisher={APS}
}

@article{carenza2024strong,
  title={Strong cosmological constraints on the neutrino magnetic moment},
  author={Carenza, Pierluca and Lucente, Giuseppe and Gerbino, Martina and Giannotti, Maurizio and Lattanzi, Massimiliano},
  journal={\prd},
  volume={110},
  number={2},
  pages={023510},
  year={2024},
  ISSN={2470-0029},
  url={http://dx.doi.org/10.1103/PhysRevD.110.023510},
  doi={10.1103/physrevd.110.023510},
  publisher={APS}
}

@article{barger2002imprint,
  title={Imprint of SNO neutral current data on the solar neutrino problem},
  author={Barger, V and Marfatia, D and Whisnant, K and Wood, BP},
  journal={Physics Letters B},
  volume={537},
  number={3-4},
  pages={179--186},
  year={2002},
  ISSN={0370-2693},
  url={http://dx.doi.org/10.1016/S0370-2693(02)01955-X},
  doi={10.1016/s0370-2693(02)01955-x},
  publisher={Elsevier}
}

@article{antia2008seismic,
  title={Seismic study of magnetic field in the solar interior},
  author={Antia, HM},
  journal={Journal of Astrophysics and Astronomy},
  volume={29},
  pages={85--92},
  year={2008},
  doi = {10.1007/s12036-008-0011-4},
  url = {https://doi.org/10.1007/s12036-008-0011-4},
  publisher={Springer}
}

@article{miranda2004constraining,
  author       = {{Miranda, O. G. and Rashba, T. I. and Rez, A. I. and Valle, J. W. F.}},
  title        = {Constraining the neutrino magnetic moment with anti-neutrinos from the Sun},
  journal      = {Phys. Rev. Lett.},
  volume       = {93},
  number       = {051304},
  year         = {2004},
  doi          = {10.1103/PhysRevLett.93.051304},
  url          = {https://doi.org/10.1103/PhysRevLett.93.051304}
}

@article{miranda2004enhanced,
  title={Enhanced solar antineutrino flux in random magnetic fields},
  author={Miranda, OG and Rashba, Timur I and Rez, AI and Valle, JWF},
  journal={\prd},
  volume={70},
  number={11},
  pages={113002},
  year={2004},
  ISSN={1550-2368},
  url={http://dx.doi.org/10.1103/PhysRevD.70.113002},
  doi={10.1103/physrevd.70.113002},
  publisher={APS}
}

@article{stock2024status,
  title={Status and Prospects of the JUNO Experiment},
  author={Stock, Matthias Raphael},
  journal={arXiv e-prints},
  year={2024},
  eprint={2405.07321},
  archivePrefix={arXiv},
  primaryClass={physics.ins-det},
  doi = {10.48550/arXiv.2405.07321},
  url={https://arxiv.org/abs/2405.07321}, 

}

@article{aliani2003determination,
  title={Determination of neutrino mixing parameters after SNO oscillation evidence},
  author={Aliani, P and Antonelli, V and Ferrari, R and Picariello, M and Torrente-Lujan, E},
  journal={\prd},
  volume={67},
  number={1},
  pages={013006},
  year={2003},
  publisher={APS},
  ISSN={1089-4918},
  url={http://dx.doi.org/10.1103/PhysRevD.67.013006},
  doi={10.1103/physrevd.67.013006},
}

@article{de2002solar,
  title={Solar neutrinos: Global analysis with day and night spectra from SNO},
  author={de Holanda, Pedro C and Smirnov, A Yu},
  journal={\prd},
  volume={66},
  number={11},
  pages={113005},
  year={2002},
  publisher={APS},
  ISSN={1089-4918},
  url={http://dx.doi.org/10.1103/PhysRevD.66.113005},
  doi={10.1103/physrevd.66.113005},
}

@article{abe2022limits,
  title={Limits on astrophysical antineutrinos with the KamLAND experiment},
  author={Abe, S and Asami, S and Gando, A and Gando, Y and Gima, T and Goto, A and Hachiya, T and Hata, K and Hayashida, S and Hosokawa, K and others},
  journal={\apj},
  volume={925},
  number={1},
  pages={14},
  year={2022},
  publisher={American Astronomical Society},
  ISSN={1538-4357},
  url={http://dx.doi.org/10.3847/1538-4357/ac32c1},
  doi={10.3847/1538-4357/ac32c1}
}

@article{abe2022search,
  title={Search for solar electron anti-neutrinos due to spin-flavor precession in the Sun with Super-Kamiokande-IV},
  author={Abe, K and Bronner, C and Hayato, Yoshinari and Ikeda, M and Imaizumi, S and Ito, Hiroshi and Kameda, J and Kataoka, Y and Miura, M and Moriyama, S and others},
  journal={Astroparticle Physics},
  volume={139},
  pages={102702},
  doi = {10.1016/j.astropartphys.2022.102702},
  year={2022},
  publisher={Elsevier},
  adsurl = {https://ui.adsabs.harvard.edu/abs/2022APh...13902702A}

}

@article{friedland2005solar,
  title={Do solar neutrinos constrain the electromagnetic properties of the neutrino?},
  author={Friedland, Alexander},  
  journal={arXiv preprint hep-ph/0505165},
  doi = {10.48550/arXiv.hep-ph/0505165},
  primaryClass = {hep-ph},
  adsurl = {https://ui.adsabs.harvard.edu/abs/2005hep.ph....5165F},
  reportNumber = "LA-UR-05-3141",
  year={2005}
}

@article{kitchatinov2008stability,
  title={Stability of toroidal magnetic fields in the radiation zone of a star},
  author={Kitchatinov, LL},
  journal={Astron. Rep},
  volume={52},
  pages={247--255},
  year={2008},
  publisher={Springer},
  doi = {10.1134/S1063772908030074},
  url = {https://doi.org/10.1134/S1063772908030074}
}

@article{PDG2024Stats,
  author        = {{Particle Data Group}},
  title         = {Statistics (Sec.~40.4): Gaussian limits and confidence intervals},
  journal       = {Prog. Theor. Exp. Phys.},
  year          = {2024},
  volume        = {2022},
  pages         = {083C01},
  doi     = {10.1093/ptep/ptac097},
  note          = {Updated Review of Particle Physics, Tables 40.1--40.2},
  url           = {https://pdg.lbl.gov/2024/reviews/rpp2024-rev-statistics.pdf}
}

@article{Abusleme_2025,
doi = {10.1088/1674-1137/ad7f3e},
url = {https://dx.doi.org/10.1088/1674-1137/ad7f3e},
year = {2025},
month = {mar},
publisher = {Chinese Physical Society and the Institute of High Energy Physics of the Chinese Academy of Sciences and the Institute of Modern Physics of the Chinese Academy of Sciences and IOP Publishing Ltd
				},
volume = {49},
number = {3},
pages = {033104},
author = {Abusleme, Angel and Adam, Thomas and Ahmad, Shakeel and Ahmed, Rizwan and Aiello, Sebastiano and Akram, Muhammad and Aleem, Abid and An, Fengpeng and An, Qi and Andronico, Giuseppe and Anfimov, Nikolay and Antonelli, Vito and Antoshkina, Tatiana and Asavapibhop, Burin and de André, João Pedro Athayde Marcondes and Auguste, Didier and Bai, Weidong and Balashov, Nikita and Baldini, Wander and Barresi, Andrea and Basilico, Davide and Baussan, Eric and Bellato, Marco and Beretta, Marco and Bergnoli, Antonio and Bick, Daniel and Bieger, Lukas and Biktemerova, Svetlana and Birkenfeld, Thilo and Morton-Blake, Iwan and Blum, David and Blyth, Simon and Bolshakova, Anastasia and Bongrand, Mathieu and Bordereau, Clément and Breton, Dominique and Brigatti, Augusto and Brugnera, Riccardo and Bruno, Riccardo and Budano, Antonio and Busto, Jose and Cabrera, Anatael and Caccianiga, Barbara and Cai, Hao and Cai, Xiao and Cai, Yanke and Cai, Zhiyan and Callier, Stéphane and Cammi, Antonio and Campeny, Agustin and Cao, Chuanya and Cao, Guofu and Cao, Jun and Caruso, Rossella and Cerna, Cédric and Cerrone, Vanessa and Chan, Chi and Chang, Jinfan and Chang, Yun and Chatrabhuti, Auttakit and Chen, Chao and Chen, Guoming and Chen, Pingping and Chen, Shaomin and Chen, Yixue and Chen, Yu and Chen, Zhangming and Chen, Zhiyuan and Chen, Zikang and Cheng, Jie and Cheng, Yaping and Cheng, Yu Chin and Chepurnov, Alexander and Chetverikov, Alexey and Chiesa, Davide and Chimenti, Pietro and Chin, Yen-Ting and Chu, Ziliang and Chukanov, Artem and Claverie, Gérard and Clementi, Catia and Clerbaux, Barbara and Molla, Marta Colomer and DiLorenzo, Selma Conforti and Coppi, Alberto and Corti, Daniele and Csakli, Simon and Dal Corso, Flavio and Dalager, Olivia and Datta, Jaydeep and De La Taille, Christophe and Deng, Zhi and Deng, Ziyan and Ding, Xiaoyu and Ding, Xuefeng and Ding, Yayun and Dirgantara, Bayu and Dittrich, Carsten and Dmitrievsky, Sergey and Dohnal, Tadeas and Dolzhikov, Dmitry and Donchenko, Georgy and Dong, Jianmeng and Doroshkevich, Evgeny and Dou, Wei and Dracos, Marcos and Druillole, Frédéric and Du, Ran and Du, Shuxian and Dugas, Katherine and Dusini, Stefano and Duyang, Hongyue and Eck, Jessica and Enqvist, Timo and Fabbri, Andrea and Fahrendholz, Ulrike and Fan, Lei and Fang, Jian and Fang, Wenxing and Fargetta, Marco and Fedoseev, Dmitry and Fei, Zhengyong and Feng, Li-Cheng and Feng, Qichun and Ferraro, Federico and Fournier, Amélie and Gan, Haonan and Gao, Feng and Garfagnini, Alberto and Gavrikov, Arsenii and Giammarchi, Marco and Giudice, Nunzio and Gonchar, Maxim and Gong, Guanghua and Gong, Hui and Gornushkin, Yuri and Göttel, Alexandre and Grassi, Marco and Gromov, Maxim and Gromov, Vasily and Gu, Minghao and Gu, Xiaofei and Gu, Yu and Guan, Mengyun and Guan, Yuduo and Guardone, Nunzio and Guo, Cong and Guo, Wanlei and Guo, Xinheng and Hagner, Caren and Han, Ran and Han, Yang and He, Miao and He, Wei and Heinz, Tobias and Hellmuth, Patrick and Heng, Yuekun and Herrera, Rafael and Hor, YuenKeung and Hou, Shaojing and Hsiung, Yee and Hu, Bei-Zhen and Hu, Hang and Hu, Jianrun and Hu, Jun and Hu, Shouyang and Hu, Tao and Hu, Yuxiang and Hu, Zhuojun and Huang, Guihong and Huang, Hanxiong and Huang, Jinhao and Huang, Junting and Huang, Kaixuan and Huang, Wenhao and Huang, Xin and Huang, Xingtao and Huang, Yongbo and Hui, Jiaqi and Huo, Lei and Huo, Wenju and Huss, Cédric and Hussain, Safeer and Imbert, Leonard and Ioannisian, Ara and Isocrate, Roberto and Jafar, Arshak and Jelmini, Beatrice and Jeria, Ignacio and Ji, Xiaolu and Jia, Huihui and Jia, Junji and Jian, Siyu and Jiang, Cailian and DiJiang, Di and Jiang, Wei and Jiang, Xiaoshan and Jing, Xiaoping and Jollet, Cécile and Kampmann, Philipp and Kang, Li and Karaparambil, Rebin and Kazarian, Narine and Ali, Khan and Khatun, Amina and Khosonthongkee, Khanchai and Korablev, Denis and Kouzakov, Konstantin and Krasnoperov, Alexey and Kuleshov, Sergey and Kutovskiy, Nikolay and Labit, Loïc and Lachenmaier, Tobias and Landini, Cecilia and Leblanc, Sébastien and Lebrin, Victor and Lefevre, Frederic and Lei, Ruiting and Leitner, Rupert and Leung, Jason and Li, Demin and Li, Fei and Li, Fule and Li, Gaosong and Li, Jiajun and Li, Mengzhao and Li, Min and Li, Nan and Li, Qingjiang and Li, Ruhui and Li, Rui and Li, Shanfeng and Li, Tao and Li, Teng and Li, Weidong and Li, Weiguo and Li, Xiaomei and Li, Xiaonan and Li, Xinglong and Li, Yi and Li, Yichen and Li, Yufeng and Li, Zhaohan and Li, Zhibing and Li, Ziyuan and Li, Zonghai and Liang, Hao and Liang, Hao and Liao, Jiajun and Limphirat, Ayut and Lin, Guey-Lin and Lin, Shengxin and Lin, Tao and Ling, Jiajie and Ling, Xin and Lippi, Ivano and Liu, Caimei and Liu, Fang and Liu, Fengcheng and Liu, Haidong and Liu, Haotian and Liu, Hongbang and Liu, Hongjuan and Liu, Hongtao and Liu, Hui and Liu, Jianglai and Liu, Jiaxi and Liu, Jinchang and Liu, Min and Liu, Qian and Liu, Qin and Liu, Runxuan and Liu, Shenghui and Liu, Shubin and Liu, Shulin and Liu, Xiaowei and Liu, Xiwen and Liu, Xuewei and Liu, Yankai and Liu, Zhen and Lokhov, Alexey and Lombardi, Paolo and Lombardo, Claudio and Loo, Kai and Lu, Chuan and Lu, Haoqi and Lu, Jingbin and Lu, Junguang and Lu, Peizhi and Lu, Shuxiang and Lubsandorzhiev, Bayarto and Lubsandorzhiev, Sultim and Ludhova, Livia and Lukanov, Arslan and Luo, Daibin and Luo, Fengjiao and Luo, Guang and Luo, Jianyi and Luo, Shu and Luo, Wuming and Luo, Xiaojie and Lyashuk, Vladimir and Ma, Bangzheng and Ma, Bing and Ma, Qiumei and Ma, Si and Ma, Xiaoyan and Ma, Xubo and Maalmi, Jihane and Magoni, Marco and Mai, Jingyu and Malyshkin, Yury and Mandujano, Roberto Carlos and Mantovani, Fabio and Mao, Xin and Mao, Yajun and Mari, Stefano M. and Marini, Filippo and Martini, Agnese and Mayer, Matthias and Mayilyan, Davit and Mednieks, Ints and Meng, Yue and Meraviglia, Anita and Meregaglia, Anselmo and Meroni, Emanuela and Meyhöfer, David and Miramonti, Lino and Mohan, Nikhil and Montuschi, Michele and Müller, Axel and Nastasi, Massimiliano and Naumov, Dmitry V. and Naumova, Elena and Navas-Nicolas, Diana and Nemchenok, Igor and Thi, Minh Thuan Nguyen and Nikolaev, Alexey and Ning, Feipeng and Ning, Zhe and Nunokawa, Hiroshi and Oberauer, Lothar and Ochoa-Ricoux, Juan Pedro and Olshevskiy, Alexander and Orestano, Domizia and Ortica, Fausto and Othegraven, Rainer and Paoloni, Alessandro and Parmeggiano, Sergio and Pei, Yatian and Pelicci, Luca and Peng, Anguo and Peng, Haiping and Peng, Yu and Peng, Zhaoyuan and Perrot, Frédéric and Petitjean, Pierre-Alexandre and Petrucci, Fabrizio and Pilarczyk, Oliver and Rico, Luis Felipe Piñeres and Popov, Artyom and Poussot, Pascal and Previtali, Ezio and Qi, Fazhi and Qi, Ming and Qi, Xiaohui and Qian, Sen and Qian, Xiaohui and Qian, Zhen and Qiao, Hao and Qin, Zhonghua and Qiu, Shoukang and Qu, Manhao and Qu, Zhenning and Ranucci, Gioacchino and Rasheed, Reem and Re, Alessandra and Rebii, Abdel and Redchuk, Mariia and Ren, Bin and Ren, Jie and Ricci, Barbara and Rientong, Komkrit and Rifai, Mariam and Roche, Mathieu and Rodphai, Narongkiat and Romani, Aldo and Roskovec, Bedřich and Ruan, Xichao and Rybnikov, Arseniy and Sadovsky, Andrey and Saggese, Paolo and Sandanayake, Deshan and Sangka, Anut and Sava, Giuseppe and Sawangwit, Utane and Schever, Michaela and Schwab, Cédric and Schweizer, Konstantin and Selyunin, Alexandr and Serafini, Andrea and Settimo, Mariangela and Sharov, Vladislav and Shaydurova, Arina and Shi, Jingyan and Shi, Yanan and Shutov, Vitaly and Sidorenkov, Andrey and Šimkovic, Fedor and Singhal, Apeksha and Sirignano, Chiara and Siripak, Jaruchit and Sisti, Monica and Smirnov, Mikhail and Smirnov, Oleg and Sogo-Bezerra, Thiago and Sokolov, Sergey and Songwadhana, Julanan and Soonthornthum, Boonrucksar and Sotnikov, Albert and Šrámek, Ondřej and Sreethawong, Warintorn and Stahl, Achim and Stanco, Luca and Stankevich, Konstantin and Steiger, Hans and Steinmann, Jochen and Sterr, Tobias and Stock, Matthias Raphael and Strati, Virginia and Studenikin, Alexander and Su, Aoqi and Su, Jun and Sun, Shifeng and Sun, Xilei and Sun, Yongjie and Sun, Yongzhao and Sun, Zhengyang and Suwonjandee, Narumon and Szelezniak, Michal and Takenaka, Akira and Tang, Jian and Tang, Qiang and Tang, Quan and Tang, Xiao and Hariharan, Vidhya Thara and Theisen, Eric and Tietzsch, Alexander and Tkachev, Igor and Tmej, Tomas and Torri, Marco Danilo Claudio and Tortorici, Francesco and Treskov, Konstantin and Triossi, Andrea and Triozzi, Riccardo and Trzaska, Wladyslaw and Tung, Yu-Chen and Tuve, Cristina and Ushakov, Nikita and Vedin, Vadim and Venettacci, Carlo and Verde, Giuseppe and Vialkov, Maxim and Viaud, Benoit and Vollbrecht, Cornelius Moritz and Sturm, Katharinavon and Vorobel, Vit and Voronin, Dmitriy and Votano, Lucia and Walker, Pablo and Wang, Caishen and Wang, Chung-Hsiang and Wang, En and Wang, Guoli and Wang, Jian and Wang, Jun and Wang, Li and Wang, Lu and Wang, Meng and Wang, Meng and Wang, Ruiguang and Wang, Siguang and Wang, Wei and Wang, Wenshuai and Wang, Xi and Wang, Xiangyue and Wang, Yangfu and Wang, Yaoguang and Wang, Yi and Wang, Yi and Wang, Yifang and Wang, Yuanqing and Wang, Yuyi and Wang, Zhe and Wang, Zheng and Wang, Zhimin and Watcharangkool, Apimook and Wei, Wei and Wei, Wei and Wei, Wenlu and Wei, Yadong and Wei, Yuehuan and Wen, Kaile and Wen, Liangjian and Weng, Jun and Wiebusch, Christopher and Wirth, Rosmarie and Wonsak, Bjoern and Wu, Diru and Wu, Qun and Wu, Yiyang and Wu, Zhi and Wurm, Michael and Wurtz, Jacques and Wysotzki, Christian and Xi, Yufei and Xia, Dongmei and Xiao, Fei and Xiao, Xiang and Xie, Xiaochuan and Xie, Yuguang and Xie, Zhangquan and Xin, Zhao and Xing, Zhizhong and Xu, Benda and Xu, Cheng and Xu, Donglian and Xu, Fanrong and Xu, Hangkun and Xu, Jilei and Xu, Jing and Xu, Meihang and Xu, Xunjie and Xu, Yin and Xu, Yu and Yan, Baojun and Yan, Qiyu and Yan, Taylor and Yan, Xiongbo and Yan, Yupeng and Yang, Changgen and Yang, Chengfeng and Yang, Jie and Yang, Lei and Yang, Xiaoyu and Yang, Yifan and Yang, Yifan and Yao, Haifeng and Ye, Jiaxuan and Ye, Mei and Ye, Ziping and Yermia, Frédéric and You, Zhengyun and Yu, Boxiang and Yu, Chiye and Yu, Chunxu and Yu, Guojun and Yu, Hongzhao and Yu, Miao and Yu, Xianghui and Yu, Zeyuan and Yu, Zezhong and Yuan, Cenxi and Yuan, Chengzhuo and Yuan, Ying and Yuan, Zhenxiong and Yue, Baobiao and Zafar, Noman and Zavadskyi, Vitalii and Zeng, Fanrui and Zeng, Shan and Zeng, Tingxuan and Zeng, Yuda and Zhan, Liang and Zhang, Aiqiang and Zhang, Bin and Zhang, Binting and Zhang, Feiyang and Zhang, Haosen and Zhang, Honghao and Zhang, Jialiang and Zhang, Jiawen and Zhang, Jie and Zhang, Jingbo and Zhang, Jinnan and Zhang, Han and Zhang, Lei and Zhang, Mohan and Zhang, Peng and Zhang, Ping and Zhang, Qingmin and Zhang, Shiqi and Zhang, Shu and Zhang, Shuihan and Zhang, Siyuan and Zhang, Tao and Zhang, Xiaomei and Zhang, Xin and Zhang, Xuantong and Zhang, Yinhong and Zhang, Yiyu and Zhang, Yongpeng and Zhang, Yu and Zhang, Yuanyuan and Zhang, Yumei and Zhang, Zhenyu and Zhang, Zhijian and Zhao, Jie and Zhao, Rong and Zhao, Runze and Zhao, Shujun and Zheng, Dongqin and Zheng, Hua and Zheng, Yangheng and Zhong, Weirong and Zhou, Jing and Zhou, Li and Zhou, Nan and Zhou, Shun and Zhou, Tong and Zhou, Xiang and Zhu, Jingsen and Zhu, Kangfu and Zhu, Kejun and Zhu, Zhihang and Zhuang, Bo and Zhuang, Honglin and Zong, Liang and Zou, Jiaheng and Züfle, Jan and (The JUNO Collaboration)},
title = {Potential to identify neutrino mass ordering with reactor antineutrinos at JUNO*},
journal = {Chinese Physics C}
}

@article{Vinyoles_2017,
doi = {10.3847/1538-4357/835/2/202},
url = {https://dx.doi.org/10.3847/1538-4357/835/2/202},
year = {2017},
month = {jan},
publisher = {The American Astronomical Society},
volume = {835},
number = {2},
pages = {202},
author = {Vinyoles, Núria and Serenelli, Aldo M. and Villante, Francesco L. and Basu, Sarbani and Bergström, Johannes and Gonzalez-Garcia, M. C. and Maltoni, Michele and Peña-Garay, Carlos and Song, Ningqiang},
title = {A New Generation of Standard Solar Models},
journal = {The Astrophysical Journal}
}

@article{PhysRevD.102.083007,
  doi = {10.1103/PhysRevD.102.083007},
  url = {https://link.aps.org/doi/10.1103/PhysRevD.102.083007},
  title = {Axion and neutrino bounds improved with new calibrations of the tip of the red-giant branch using geometric distance determinations},
  author = {Capozzi, Francesco and Raffelt, Georg},
  journal = {Phys. Rev. D},
  volume = {102},
  issue = {8},
  pages = {083007},
  numpages = {14},
  year = {2020},
  month = {Oct},
  publisher = {American Physical Society}
}

@article{Heger_2009,
doi = {10.1088/0004-637X/696/1/608},
url = {https://doi.org/10.1088/0004-637X/696/1/608},
year = {2009},
month = {apr},
publisher = {The American Astronomical Society},
volume = {696},
number = {1},
pages = {608},
author = {Heger, Alexander and Friedland, Alexander and Giannotti, Maurizio and Cirigliano, Vincenzo},
title = {THE IMPACT OF NEUTRINO MAGNETIC MOMENTS ON THE EVOLUTION OF MASSIVE STARS},
journal = {The Astrophysical Journal}
}

@article{lishaoping,
    author = {Li, Shao-Ping and Xu, Xun-Jie},
    title = {Neutrino magnetic moments meet precision Neff measurements},
    journal = {Journal of High Energy Physics},
    year = {2023},
    volumen = {2023},
    number = {2},
    pages = {85},
    date = {2023/02/08},
    doi = {10.1007/JHEP02(2023)085},
    id = {Li2023},
    isbn = {1029-8479},
    url = {https://doi.org/10.1007/JHEP02(2023)085}
}

@article{Huber:2011wv,
    author = "Huber, Patrick",
    title = "{On the determination of anti-neutrino spectra from nuclear reactors}",
    eprint = "1106.0687",
    archivePrefix = "arXiv",
    primaryClass = "hep-ph",
    doi = "10.1103/PhysRevC.85.029901",
    journal = "Phys. Rev. C",
    volume = "84",
    pages = "024617",
    year = "2011",
    note = "[Erratum: Phys.Rev.C 85, 029901 (2012)]"
}

@article{Mueller.83.054615,
  title = {Improved predictions of reactor antineutrino spectra},
  author = {Mueller, Th. A. and Lhuillier, D. and Fallot, M. and Letourneau, A. and Cormon, S. and Fechner, M. and Giot, L. and Lasserre, T. and Martino, J. and Mention, G. and Porta, A. and Yermia, F.},
  journal = {Phys. Rev. C},
  volume = {83},
  issue = {5},
  pages = {054615},
  numpages = {17},
  year = {2011},
  month = {May},
  publisher = {American Physical Society},
  doi = {10.1103/PhysRevC.83.054615},
  url = {https://link.aps.org/doi/10.1103/PhysRevC.83.054615}
}

@software{Ventura2025Software,
  author       = {Ventura and Saul Panibra},
  title        = {JUNO Event-Rate Calculator and Background Extrapolation Framework},
  year         = {2025},
  publisher    = {Zenodo},
  version      = {v1.0},
  doi          = {10.5281/zenodo.17613795},
  url          = {https://doi.org/10.5281/zenodo.17613794}
}
\end{CJK*}

\end{document}